\documentclass[10pt,aps,prb,superscriptaddress,twocolumn,showpacs]{revtex4-1}

\usepackage[english]{babel}
\usepackage[colorlinks=true,urlcolor=blue,citecolor=blue,linkcolor=blue]{hyperref}
\usepackage{times}
\usepackage{amsmath,amssymb}
\usepackage[pdftex]{graphicx}
\usepackage{url}

\usepackage{algpseudocode}

\newcommand{\be}{\begin{equation}}
\newcommand{\ee}{\end{equation}}

\graphicspath{{figures/}}

\begin{document}

\title{From local to global ground states in Ising spin glasses}

\author{Ilia Zintchenko}
\affiliation{Theoretische Physik, ETH Zurich, 8093 Zurich, Switzerland}

\author{Matthew B. Hastings}

\affiliation{Station Q, Microsoft Research, Santa Barbara, CA
  93106-6105, USA}

\affiliation{Quantum Architectures and Computation Group, Microsoft
  Research, Redmond, WA 98052, USA}

\author{Matthias Troyer}
\affiliation{Theoretische Physik, ETH Zurich, 8093 Zurich, Switzerland}
\begin{abstract}
  We consider whether it is possible to find ground states of
  frustrated spin systems by solving them locally. Using spin glass
  physics and Imry-Ma arguments in addition to numerical benchmarks we
  quantify the power of such local solution methods and show that for
  the average low-dimensional spin glass problem outside the
  spin-glass phase the exact ground state can be found in polynomial
  time. In the second part we present a heuristic, general-purpose
  hierarchical approach which for spin glasses on chimera graphs and
  lattices in two and three dimensions outperforms, to our knowledge,
  any other solver currently around, with significantly better scaling
  performance than simulated annealing.
\end{abstract}

\maketitle

\section{Introduction}

The combination of disorder and frustration in spin glasses
\cite{BinderYoung} creates a complex energy landscape with many local
minima that makes finding their ground states a formidable
challenge. In particular finding the assignments of spins $s_i=\pm1$
which minimises the total energy of an Ising spin glass with
Hamiltonian \be \label{Hdef} H = \sum_{ij}J_{ij}s_i s_j + \sum_{i} h_i
s_i, \ee where $s_i = \pm 1$ and $J_{ij},h_i \in \mathbb{R}$, is
non-deterministic polynomial (NP) hard \cite{Barahona1982} and no
polynomial time algorithm is known for the hardest
instances. NP-hardness also means that any problem in the complexity
class NP can be mapped to an Ising spin glass with only polynomial
overhead. This includes the travelling salesman problem, satisfiability
of logical formulas, and many other hard optimization
problems. Explicit mapping for a number of these problems have
recently been given in Ref. \onlinecite{Lucas2014}. Efficient solvers
for Ising spin glass problems hence can have an impact far beyond spin
glass physics.

This broad spectrum of applications has also motivated the development
of the devices by the Canadian company D-Wave Systems
\cite{Harris2010,0953-2048-23-6-065004,berkley2010scalable,Johnson2011}. These
devices have been designed to employ quantum annealing
\cite{Kadowaki1998} for Ising spin glass problems using
superconducting flux qubits. However, it has not yet been shown that
they can outperform classical devices \cite{SSSV,Ronnow2014}.
Determining the complexity of solving the spin glass problems on the
so-called ``chimera graph'', which is implemented by the hardware of
the D-Wave devices, and finding the best classical algorithms for them
is important in the search for quantum speedup on these devices
\cite{Ronnow2014}.

Motivated by these comparisons and the importance of efficiently
solving Ising spin glass problems, here we consider the complexity of
solving such problems for random spin glass instances on
finite-dimensional lattices, including the chimera graph. In
Sec. \ref{sec:boundary} we discuss the effects of non-zero temperature
and magnetic field on Ising spin glasses and argue that the absence of
correlations outside the spin glass phase allows for polynomial time
algorithms. Section \ref{sec:exact} presents an exact solver based on
this idea which solves the system quasi-locally by considering finite
patches of the lattice. Finally, in Sec. \ref{sec:hierarchical} we
present a hierarchical heuristic approach, which recursively solves
groups of spins by splitting each group into smaller sub-groups. For
our benchmark problems on two and three dimensional periodic lattices
and chimera graphs with random disorder this approach outperforms, to
our knowledge, any other solver currently available and scales
significantly better than simulated annealing. While we give a
qualitative explanation of the advantage of the hierarchical solver,
it remains an open theoretical question to give a quantitative
argument for its improved scaling. The interested reader can skip
directly to this section as it can be understood independently of the
scaling analysis earlier in the paper.

\section{Boundary Condition Dependence in Frustrated Spin Systems}
\label{sec:boundary}

It is evident that if the fields $h_i$ in Eq. (\ref{Hdef}) are very
large, the problem can be solved by simply aligning each spin relative
to the field.  The problem becomes more difficult at smaller $h_i$,
and the meaningful question is whether a phase transition intervenes
at some non-zero value of the field strength, where the difficulty
increases greatly. In this section, we argue that the relevant
transition indeed is already known in the literature, where it is
referred to as the de Almeida-Thouless line.  We argue that above this
transition (which happens for {\it any} non-zero random choice of
$h_i,J_{ij}$ in two dimensions), the problem can be solved by
considering larger patches of spins, with the patch size diverging as
the field strength goes to zero; the spins in the middle of these
large patches become independent of those outside the patch and can be
fixed using a local algorithm. We first review the relevant literature
at $h_i=0$ and the scaling theory at small $h_i$.

\subsection{Review}
In the particular ensemble where fields vanish ($h_i=0$), the behaviour
of the model depends strongly upon both the dimensionality of the
system and upon the choice of the ensemble for the couplings between
spins. In this discussion, we will focus on the case of a continuous
distribution, e.g. a Gaussian one, with vanishing mean.

We will also refer to results in the literature that study
nearest-neighbour couplings on a square or cubic lattice, rather than
the chimera graph. One important distinction between the
two-dimensional square lattice and the chimera graph is that for the
square lattice, as for any planar graph, if the magnetic fields vanish
there are efficient polynomial time matching algorithms for finding
exact ground states \cite{matching}, while on non-planar graphs, such
as the chimera, it is NP-hard. We discuss this further below.

In two dimensions it is accepted that there is no spin glass phase at
temperature $T>0$ \cite{Hartmann01}.  To quantify this, consider a
pair of sites $i,j$.  Let $\langle \ldots \rangle$ denote the thermal
average of an operator at temperature $T$ and let $[\ldots]_H$ denote
the disorder average over Hamiltonians $H$.  Since the couplings are
chosen with zero mean, we have that $[\langle s_i s_j \rangle]_H=0$
exactly.  However, generically the ground state is unique and hence
$[(\langle s_i s_j \rangle)^2]_H=1$ at $T=0$, and this average is
expected to be positive at $T>0$; however, the average vanishes in the
limit of large distances between $i,j$.

The reason for the absence of a spin glass phase is that it costs very
little energy to flip a domain of spins.  Consider flipping a cluster
of spins of linear size $\ell$.  In a ferromagnetic state, this costs
energy proportional to $\ell$.  In a spin glass ground state, it is
possible, however, that a cluster can be found which costs very low
energy to flip.  Using various methods of generating flipped patches
(by for example boundary condition changes), it is found that the
energy of the domain wall scales proportional to $\ell^{\theta}$ with
$\theta \approx -0 .282(2)$ \cite{Hartmann01}.  Thus, it costs {\it
  less} energy to flip larger clusters, and no matter how small $T$
is, for $T>0$ there eventually will be some $\ell$ such that flipping
clusters at that scale costs energy smaller than $T$.  Hence there
will be many thermally excited domain walls.  On the other hand, for
three dimensions and higher, there is believed to be a phase
transition temperature $T_c>0$ with a domain wall exponent $\theta>0$
for excitations above the ground state \cite{mcmillan}.

Similarly, we can consider random models with nonzero fields
\cite{field0,field1,field2} and denote standard deviation of the field
magnitude by $h$ . In this case, we consider the quantity
$$\left[\Bigl( \langle s_i s_j \rangle - \langle s_i \rangle \langle
  s_j \rangle \Bigr)^2\right]_H$$.  If this quantity tends to a
non-zero limit at large distance between $i,j$, then we term this a
spin glass phase.  It has been shown that such a spin-glass phase can
exist in a mean-field model at $h\neq 0$ \cite{at}; the line in the
$h-T$ plane separating the spin glass from the paramagnetic phase is
termed the de Almeida-Thouless line.  However, it is unclear whether
such a spin glass phase at $h \neq 0$ can persist in a local
finite-dimensional model.  Numerical work \cite{katzgraber09,larson13}
suggests that it exists for dimension $d>d_c=6$.  However, it is
accepted that the spin glass phase at $h\neq 0$ does not persist in
dimension $d=2$ and in the next subsection we will explain why this is
expected given the exponent $\theta$ discussed above.

It should be emphasised that it is not necessarily difficult to find
ground states in a spin glass phase, as exemplified by the matching
algorithm for the planar case in $d=2$ at $h=0$. Conversely, even if a
random ensemble is not in the spin glass phase, particular instances
may be difficult, as exemplified by the fact that in $d=2$ at $h \neq
0$ the model is not in a spin glass phase, but finding the ground
state of arbitrary instances is still NP-hard.

\subsection{Weak Field Scaling in $d=2$}
\label{sub_sec:domain_wall}
We now consider the effect of a weak magnetic field $h\neq 0$ in
$d=2$.  Our general goal is to show that in this case, we expect that
the value of a given spin in the ground state can often be fixed using
a purely {\it local} calculation.  The argument is a version of the
Imry-Ma argument applied to disordered systems \cite{imryma} and in the
specific application to spin glasses is an example of the droplet
picture \cite{droplet}.  We conjecture that a similar argument (with
different exponents) will work if there is no de Almeida-Thouless line (i.e., whenever
there is no spin glass phase at non-zero magnetic field).

Consider a spin $s_\text{cent}$ at the center of a patch of size $\ell$ inside a
larger system of linear size $L$. Suppose that we have found some
configuration of spins which is a ground state.  At $h=0$, it is
impossible to know whether $s_\text{cent}=+1$ or $s_\text{cent}=-1$
without knowing the value of the boundary spins because there is a
$Z_2$ symmetry.  However, at $h \neq 0$, it may be possible to
determine the value of the spin $s_i$ independent of the value of the
boundary spins.  That is, there may be some choice (either
$s_\text{cent}=+1$ or $s_\text{cent}=-1$) that minimises the energy
inside the patch for {\it all} choices of boundary spins.  In this
case, we know that in the global ground state the spin $s_\text{cent}$
will take the given value.

\begin{figure}
  \centering
  \includegraphics[width=0.7\columnwidth]{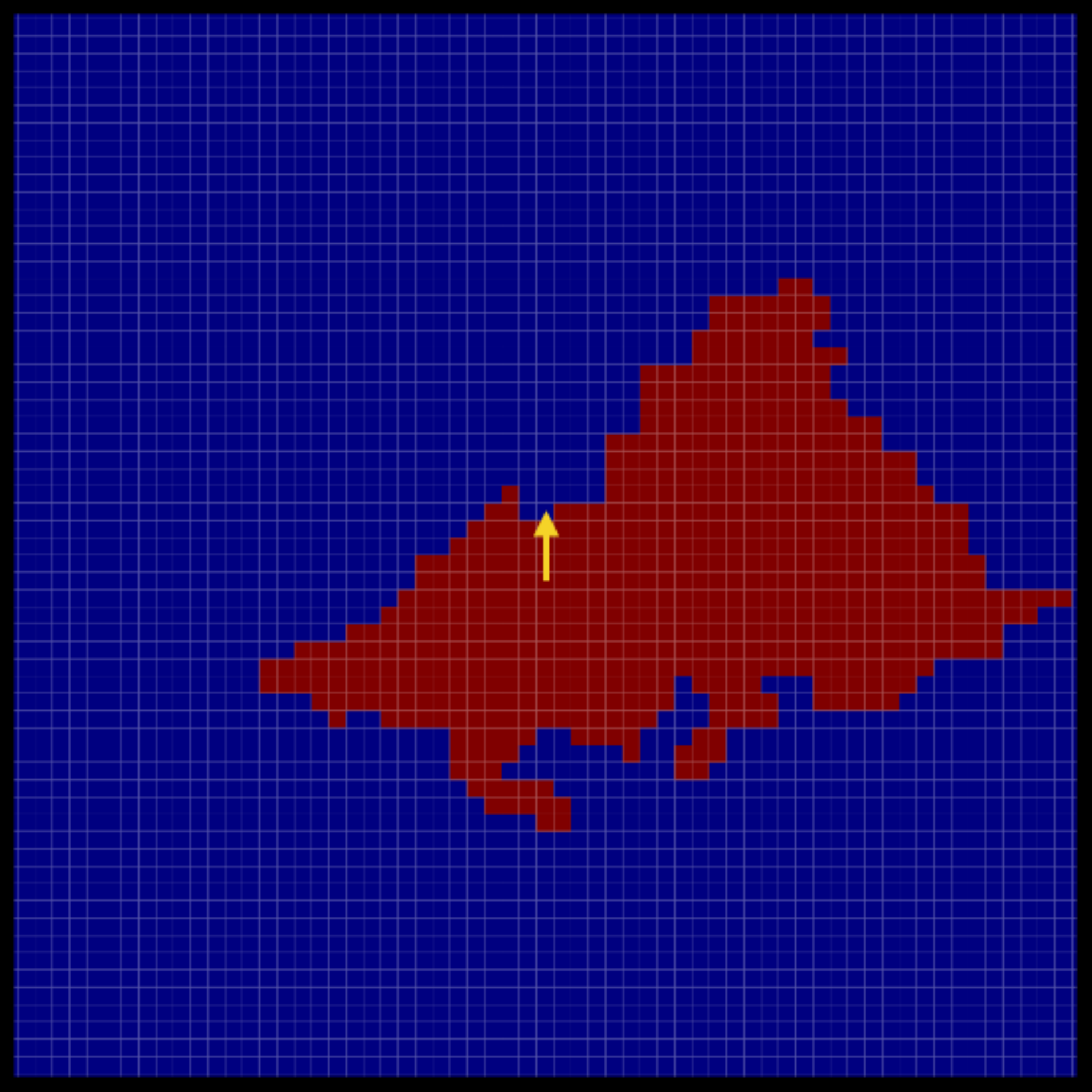}
  \caption{If central spin, is forced to be opposite to its optimal
    orientation while keeping the spins on the boundary of the patch
    fixed, a cluster of spins around it will also flip. Central spin
    marked in \emph{yellow}, flipped cluster marked in \emph{red} and
    the boundary in \emph{black}}
\label{fig:flipped}
\end{figure}

To analyse the ability to fix the spin independently of boundary
conditions, we again begin with the case $h=0$ to develop a scaling
argument that will apply at small $h$.  Consider a given configuration
of boundary spins, which we write as $\vec s_\text{bdry}$, where we
write this as a vector to emphasise that there are many boundary
sites.  At $h=0$, we can minimise the energy inside the patch for this
choice of boundary spins, uniquely fixing all spins inside the patch.
Suppose that this minimisation gives $s_\text{cent}=+1$.  Now consider
the case in which we force $s_\text{cent}=-1$, defining a new
configuration of spins inside the patch which minimises the energy
subject to the given boundary conditions $\vec s_\text{bdry}$ and
given that $s_\text{cent}=-1$.  Forcing $s_\text{cent}$ to take the
opposite value will flip also a cluster of spins around the central
spin, creating a domain wall around that cluster of spins, as shown in
Fig.~\ref{fig:flipped}.  The energy of this domain wall will be
proportional to $\ell^{\theta}$ which therefore decreases with increasing
$\ell$.

\begin{figure}
  \centering
  \includegraphics[width=1.0\columnwidth]{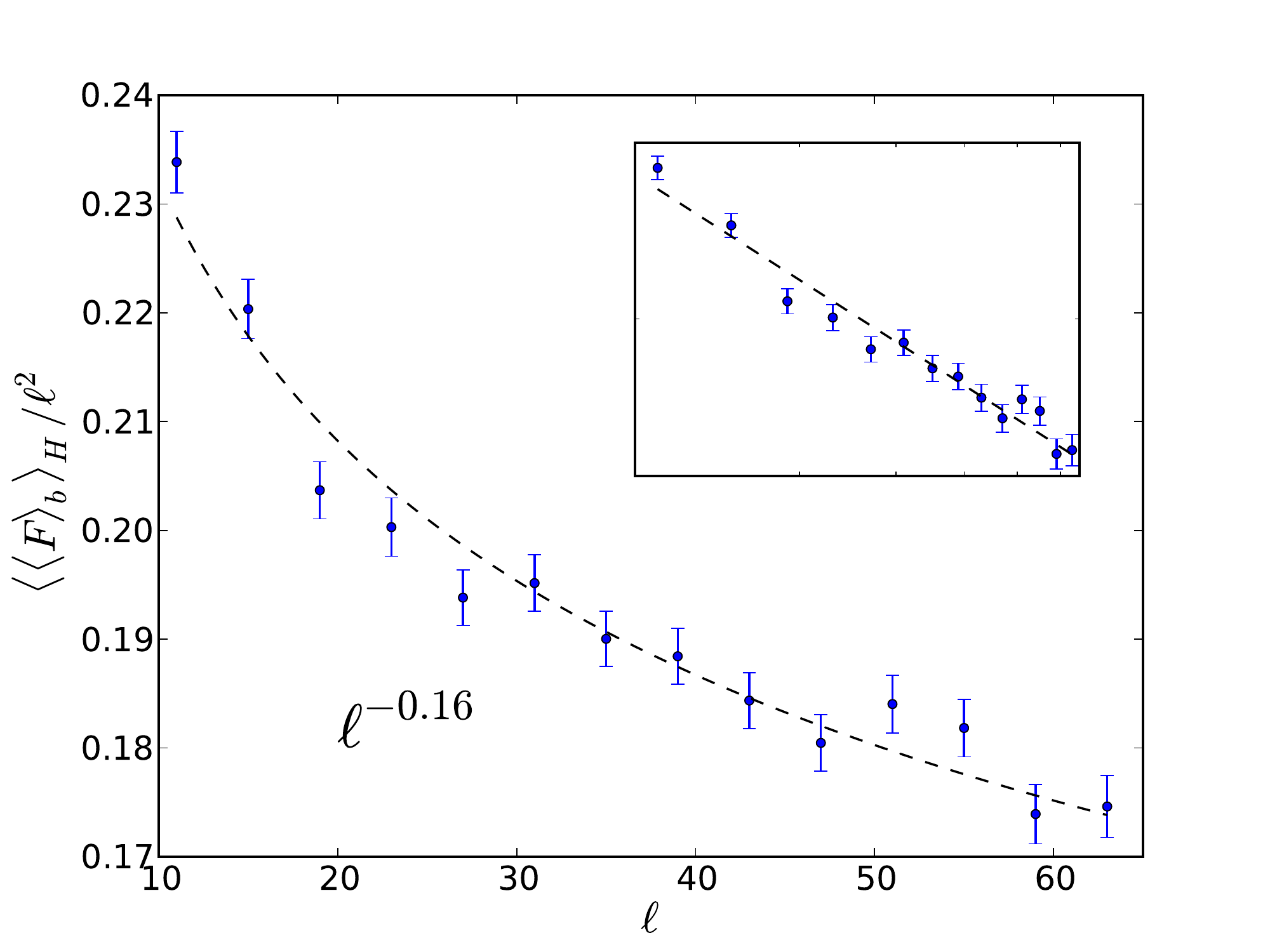}
  \caption{Number of spins in the cluster $F$ averaged over different
    boundary configurations and Hamiltonians vs. linear dimension of
    patch $\ell$. The data is fitted with to a power law (dashed
    line). Inset show log-log scale}
\label{fig:hammingdistance}
\end{figure}

The number of spins in the cluster scales also as a power of $\ell$,
with the power slightly less than \cite{Kawashima99} $2$; in that
reference, the exponent $1.80(2)$ was found for one specific method of
constructing droplets.  Our numerical studies, shown in
Fig.~\ref{fig:hammingdistance}, indicate that the number scales as
$\ell^{d_\text{clust}}$, with a fractal dimension
$d_\text{clust}\approx 1.84$; while this dimension might revert to $2$
for larger system sizes, we use the fractal dimension extracted at
these system sizes to facilitate comparison with our complexity
analysis below.

This cluster then defines a larger effective spin.  The cost to flip
this effective spin relative to the rest of the patch is proportional
to $\ell^{\theta}$.  We now consider the case that $h\neq 0$, and analyse
the effect of the non-zero $h$ on this effective spin.  Given that the
magnetic fields acting on the spins in the cluster are chosen
randomly, we expect that the cluster will experience an effective
magnetic field \be h_{eff} \propto h \ell^{d_\text{clust}/2}.  \ee
Balancing these energy scales, we find that \be h_{eff} \sim
\ell^{\theta} \rightarrow \ell \propto
h^{-\frac{1}{-\theta+d_\text{clust}/2}}.  \ee

\begin{figure}
  \centering
  \includegraphics[width=1.0\columnwidth]{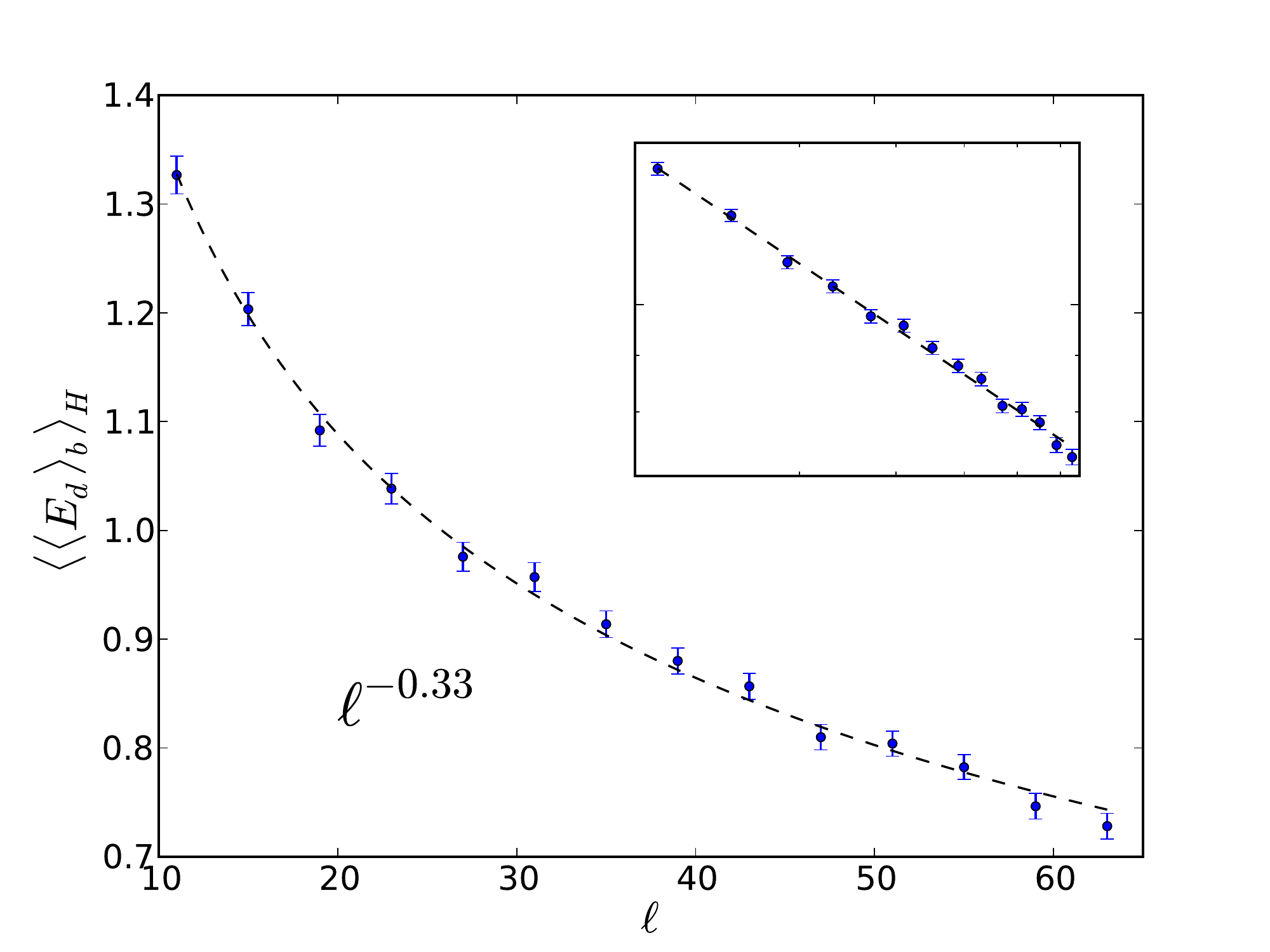}
  \caption{Defect energy $E_d$ averaged over different boundary
    configurations and Hamiltonians vs. linear dimension of patch
    $\ell$. The data is fitted with to a power law (dashed
    line). Inset show log-log scale}
\label{fig:defectenergy}
\end{figure}

In Fig.~\ref{fig:defectenergy}, we show our estimate for $\theta
\approx -0.33$ obtained from the defect energy $E_d$ gained when the
central spin is forced in opposite direction to its optimal with a
fixed boundary configuration around a patch. While this differs
slightly from the numbers quoted above, we remark that many different
ways of forcing domain walls in have been considered in the
literature, such as flipping a central spin as here or changing global
boundary conditions and these may give rise to different values,
especially for finite sizes; see
Refs.~\onlinecite{Kawashima00,Kawashima99,hartmann04} for various
possibilities.  Thus, we get that \be \label{scalingLh} \ell \propto
h^{-0.8}.  \ee For $\ell$ larger than this number, the coupling of the
cluster to the effective field exceeds its coupling to the rest of the
patch, so that the value of the central spin can be fixed {\it
  independently} of boundary conditions.  Note that this analysis
focuses on one possible way to fix in which the central spin can
become independent of boundary conditions; others may be possible.

\subsection{Boundary Condition Dependence}
The above scaling analysis gives an estimate of the length scale at
which we can fix the central spin in a patch.  The total number of
spins which can be fixed in the system depends on the local fields and
patch size and can be estimated from the probability of fixing a
single spin. To quantify this probability we define \be \chi_B(h,\ell) =
1-\left[ \Bigl( [ s_c ]_B \Bigr)^2 \right]_H \ee where $s_c$ is the
central spin and where $[\ldots]_B$ denotes the average over boundary
conditions.  We term this quantity $\chi_B$ as it measures the
response of the central spin to change in boundary conditions.  If
this quantity is equal to $0$, then the spin can be fixed
independently of boundary conditions as it assumes the same value for
all choice of boundary.

\begin{figure}
  \includegraphics[width=1.0\columnwidth]{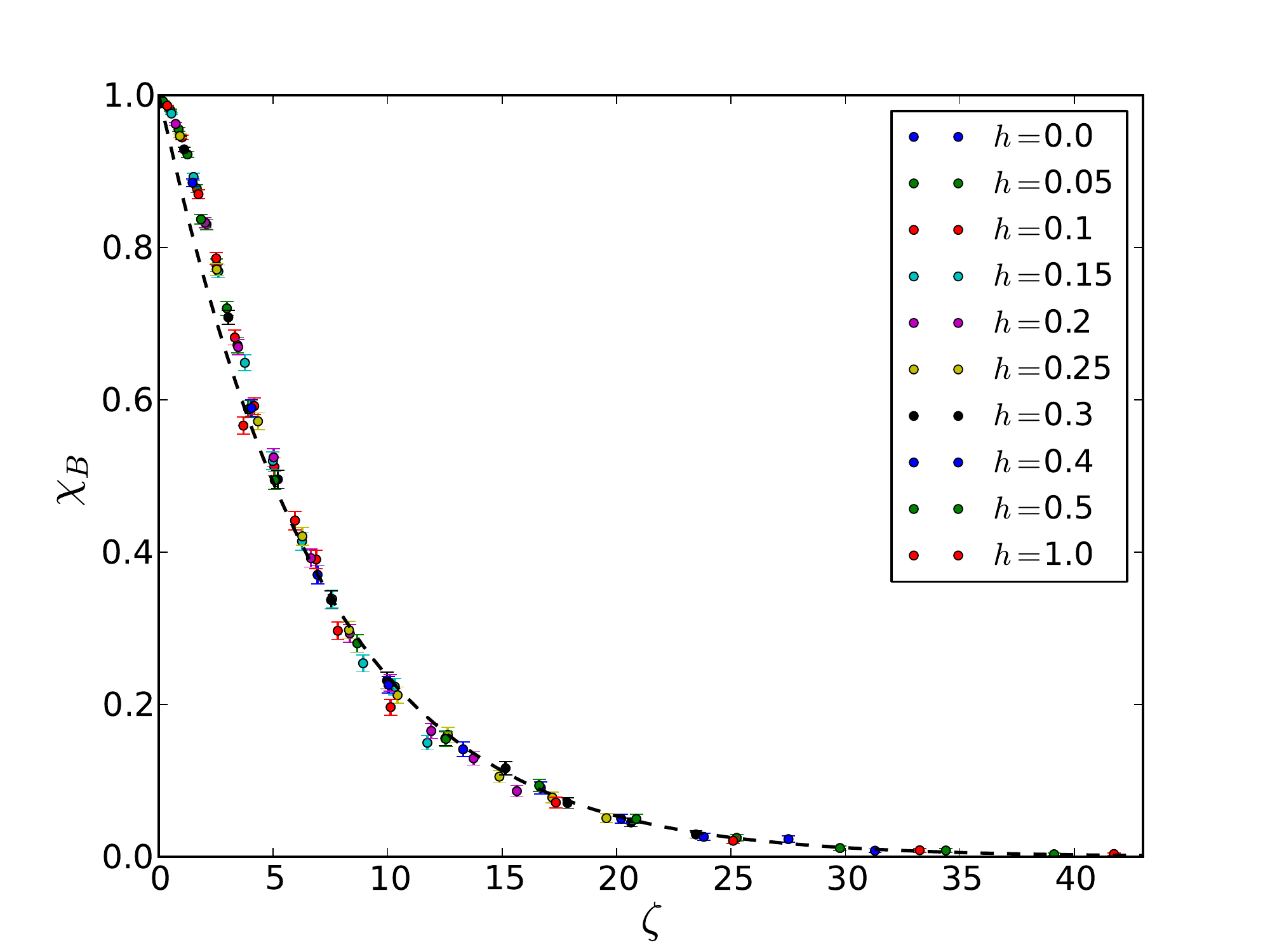}
  \caption{Correlation of central spin with the boundary for different
    fields and patch sizes. $\zeta = h \cdot \ell^{1.19}$. The fit of
    $\chi_B(h,\ell)$ is to $2^{-a \cdot \zeta^b}$, where $a = 0.198$ and
    $b = 1.02$ (dashed line)}
  \label{fig:spin_correlation}
\end{figure}

For this averaged quantity, we find a scaling collapse as shown in
Fig.~\ref{fig:spin_correlation}.  The scaling collapse onto a single
curve is implement by defining the scaling variable $\xi=h \cdot
\ell^{1.19}$.  This implies a scaling
$$\ell \sim h^{1/1.19}=h^{0.840\ldots}$$.  This should be compared with the
estimate in Eq.~(\ref{scalingLh}); the agreement of exponents is
reasonable, and if we use $\theta=-0.28$ instead of our measured
$\theta=-0.33$ the agreement becomes more accurate.  We find that the
scaling collapse can be approximately fit by the form \be \chi_B(h,\ell) =
\exp{(-\text{poly}(h,\ell))}.  \ee

To obtain statistical information about whether we can fix a spin
independently of boundary, it suffices to determine the behaviour of
$\chi_B$ in the tail, see Fig. \ref{fig:spin_correlation}, where we
fit $\chi_B = 2^{-a \cdot \zeta^b}$ with the constants $a$ and $b$. We
cannot be completely confident about the tail behaviour of $\chi_B$ at
large $h,\ell$ from these simulations, but let us use this estimate to
try to determine the complexity of a simple solver which tries to
solve each spin by taking a sufficiently large patch that $\chi_B=0$.
The complexity of the solver will depend upon the scaling of $\chi_B$,
but we will estimate that it takes a polynomial time (in $N$) for any
non-zero $h$. We will find in the next section that we can improve on
this, by using the fact that once a single spin is fixed it simplifies
the fixing of other spins.

Since there are only $2^{4\ell}$ possible boundary conditions, the
minimum non-zero value of $\chi_B$ is of order
$2^{-4\ell}$. Considering the $N$ possible choices of central spin,
only $\mathcal{O}(1)$ spins correlate with the boundary if
$\exp{(-\text{poly}(h,\ell))} = \mathcal{O}(1/N) 2^{-4\ell}$.
Equivalently, this holds if $2^{4\ell} \exp{(-\text{poly}(h,\ell))} =
\mathcal{O}(1/N)$; since $\zeta^b>1$, the scaling of the left-hand
side of this equation is dominated by the second term. Hence, the
equation will hold when \be h \cdot \ell^{1.19} \sim \log({N})^{1/b}.
\ee Thus, we expect that for $\ell$ larger than this, it will be
possible to fix all spins.

Since each patch can be solved exactly with complexity
$\exp\left\{\ell\right\}$ using a dynamic programming method
\cite{treewidth}, at a fixed $h$ the whole system can be solved with
complexity \be {\rm poly}(L) \exp\{h^{-1/1.19} \cdot (\log N)^{1/(1.19
  \cdot b)}\}.  \ee Since $b>1$ and $1/(1.19 \cdot b) < 1$, the
exponential term is sub-linear and the total complexity is therefore
polynomial; however, it diverges as $h \rightarrow 0$.  The exact
estimate may depend sensitively upon the tail of the curve which we
cannot determine with full confidence.

It should, however, be emphasised that the data in
Fig.~\ref{fig:spin_correlation} arises only from an average over a
finite number (in this case, $1000$) of boundary conditions.  This
finite number was chosen to enable rapid sampling of the curve.  To
exactly solve a specific sample, we need to consider {\it all}
possible boundary conditions, as discussed in the next section.

\section{Finding the exact global ground state}
\label{sec:exact}

Following the argument above, correlations in a typical
finite-dimensional lattice decay exponentially if $h > 0$ and the
ground state for such a system can therefore be found in polynomial
time as the optimal orientation of single spins can be determined with
high probability by considering only finite regions of the
system. Furthermore, even for zero fields we present strong numerical
evidence that the typical two-dimensional case can be solved in
polynomial time with a more general approach which we describe below.

\subsection{Single spin reduction}

Let us consider a spin in the center of a patch in our system. If for
all boundary configurations of the patch the optimal orientation of
the central spin is the same, then it is independent of the boundary
and can thus be fixed to that value. Based on this idea, the simplest
way to find the ground state is by determining the optimal orientation
of each spin independently by building a patch around it and checking
if the optimal orientation of the central spin is independent of the
boundary. If this is not the case, we increase the patch size and
check again until the spin becomes independent of the boundary. When
all spins are fixed the system is solved.

This approach can be further improved by solving the system similar to
a crossword puzzle rather than considering each spin independently. If
a spin gets fixed, this will reduce the number of possible
configurations for patches containing that spin, which in return may
allow more spins to get fixed without increasing the patch sizes.

Fixing single spins is a simple algorithm which can be very efficient
for systems with large fields. In the limit of very large fields the
complexity approaches $\mathcal{O}(N)$ as each spin becomes
independent of its neighbours. However, for small fields the
computational effort increases as the correlation length diverges when
the field approaches zero requiring patches comparable to the total
system size. A more general approach, discussed next, remains
effective in that limit.

\subsection{Patch reduction}

\begin{figure}
  \includegraphics[width=0.6\columnwidth]{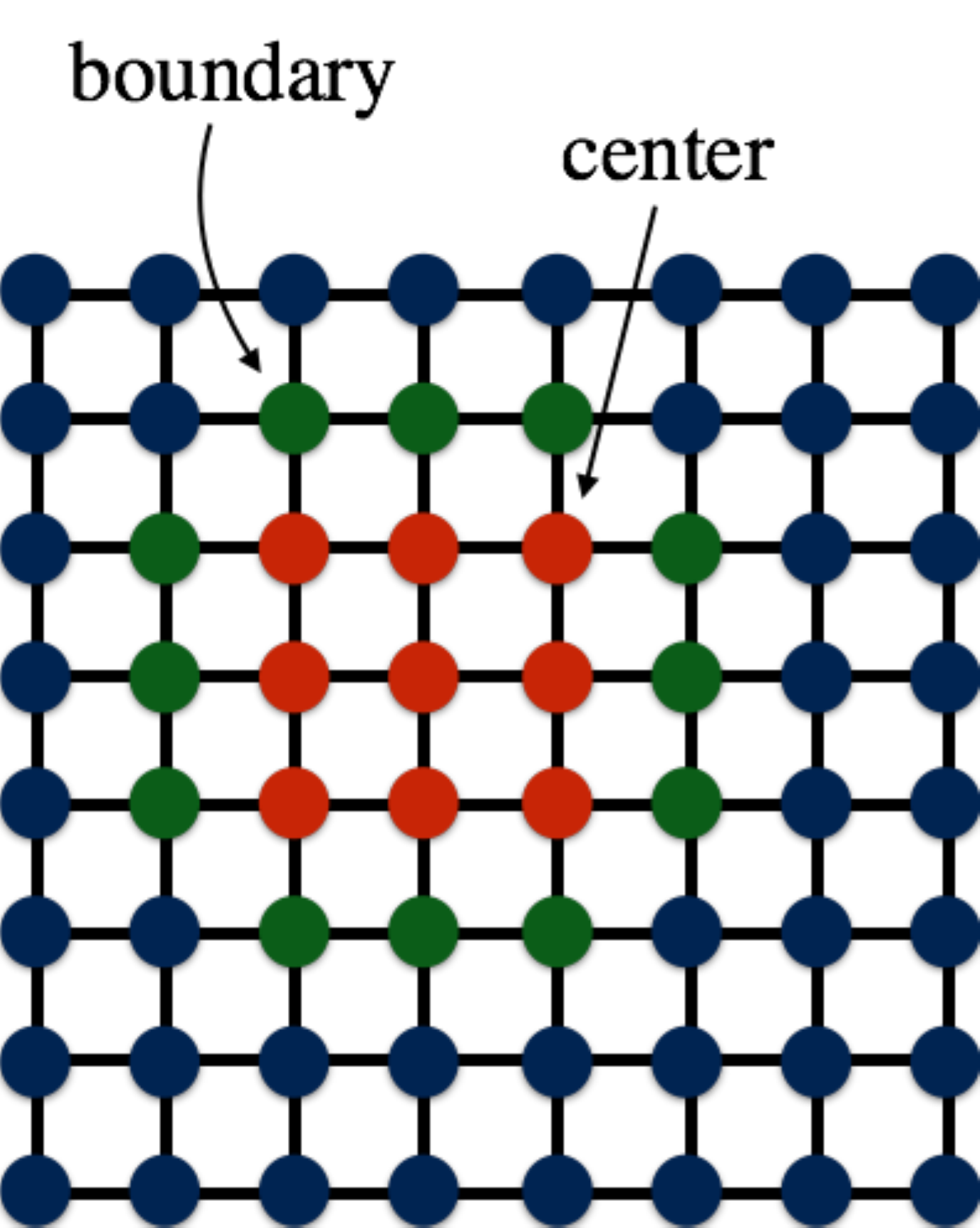}
  \caption{Illustration of a patch. Central spins are marked in
    \emph{red}, boundary spins in \emph{green} and the rest of the
    spins in the system are \emph{blue}.}
  \label{fig:center_boundary}
\end{figure}

Instead of only attempting to fix the central spin, correlations
between spins inside a patch can be captured by considering all
possible configurations of a patch that minimise the energy for a
given choice of boundary conditions (see
Fig.~\ref{fig:center_boundary}). These configurations are then
constrained by requiring consistency between overlapping patches. We
find numerically that this approach is significantly more efficient
than the single spin algorithm.

The algorithm starts with a small patch size (e.g. a single spin in
the center) and sequentially builds patches around each spin. For each
boundary configuration of a given patch we store the configuration of
the boundary together with the corresponding optimal configuration of
the center spins.  If the local ground state of a patch turns out to
be degenerate for a given boundary condition, we arbitrarily pick any
of these configurations if our aim is to obtain just one of the
potentially degenerate global ground states. Note that if instead we
are interested in finding all ground states, then for each boundary
configuration all degenerate interior configurations need to be
stored.

The number of potential ground state configurations within a patch
(boundary and interior) is then further reduced by removing those
configurations which are inconsistent with the constraints imposed by
overlapping patches.

After a pass through all spins we increase the patch size and repeat
the above steps with larger patches until only a single configuration
remains or all remaining configurations have the same energy.  As the
patch size increases, the set of configurations which satisfy all
constraints is strongly reduced and typically scales much better than
the exponential worst case.

\subsection{Improved patch reduction}

One way to significantly reduce the cost of storing configurations is
by removing some spins from the system. If for a pair of neighbouring
spins $s_i$ and $s_j$, their product $s_i s_j$ is constant in all
configurations, they can be replaced by a single spin. If only one
ground state is targeted, this procedure will finally eliminate all,
but one spin. More generally, any arbitrary spin can be removed by
replacing it with multi-spin interactions such that for each
configuration of the neighbouring spins the local energy is conserved
given that the spin to be removed aligns optimally with respect to its
neighbours.

\subsection{Empirical scaling}

\begin{figure}
  \includegraphics[width=1.0\columnwidth]{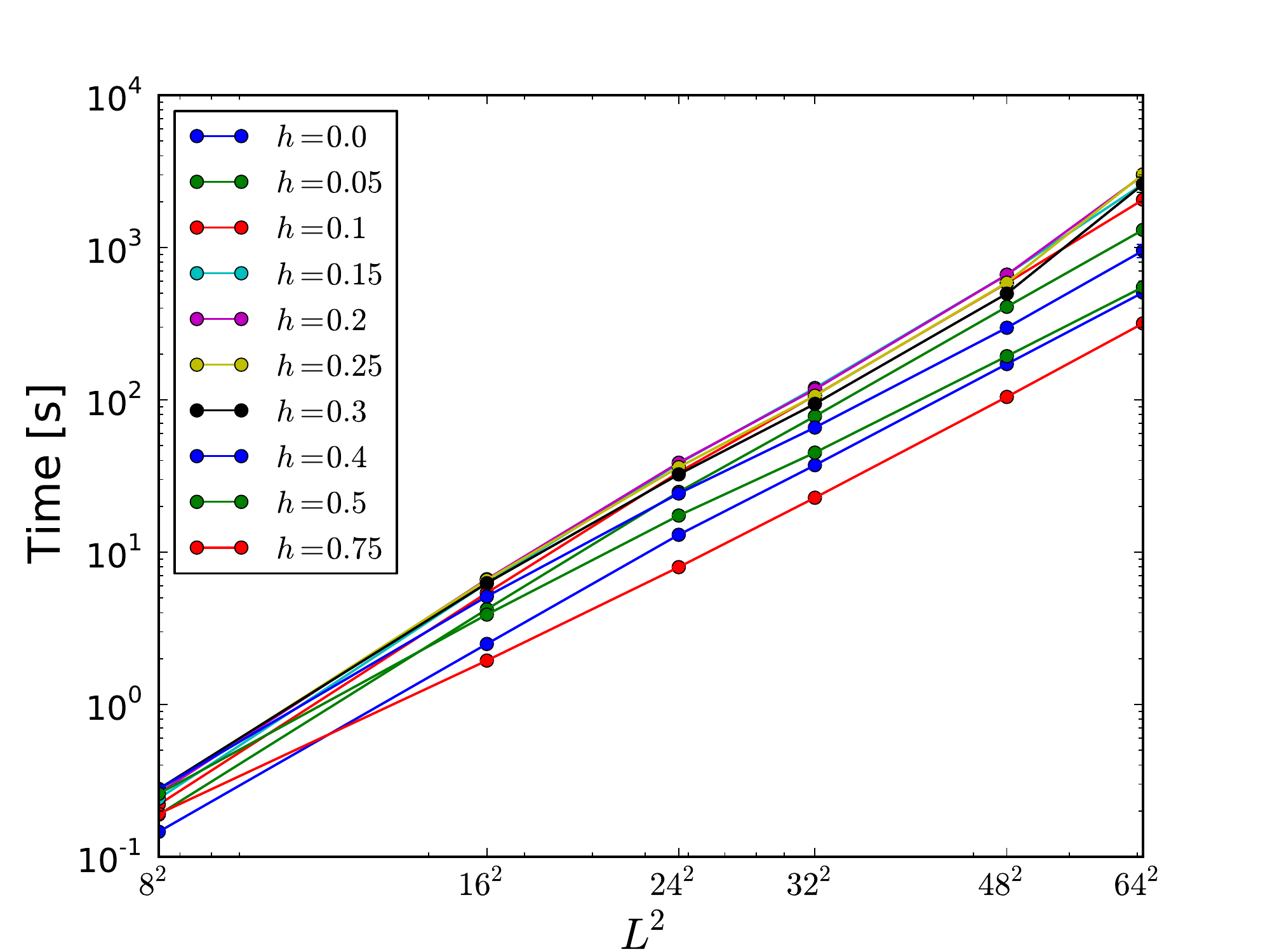}
  \caption{Median wall clock time (in seconds) for different system
    sizes and various fields}
  \label{fig:complexity_L_runtime}
\end{figure}

\begin{figure}
  \includegraphics[width=1.0\columnwidth]{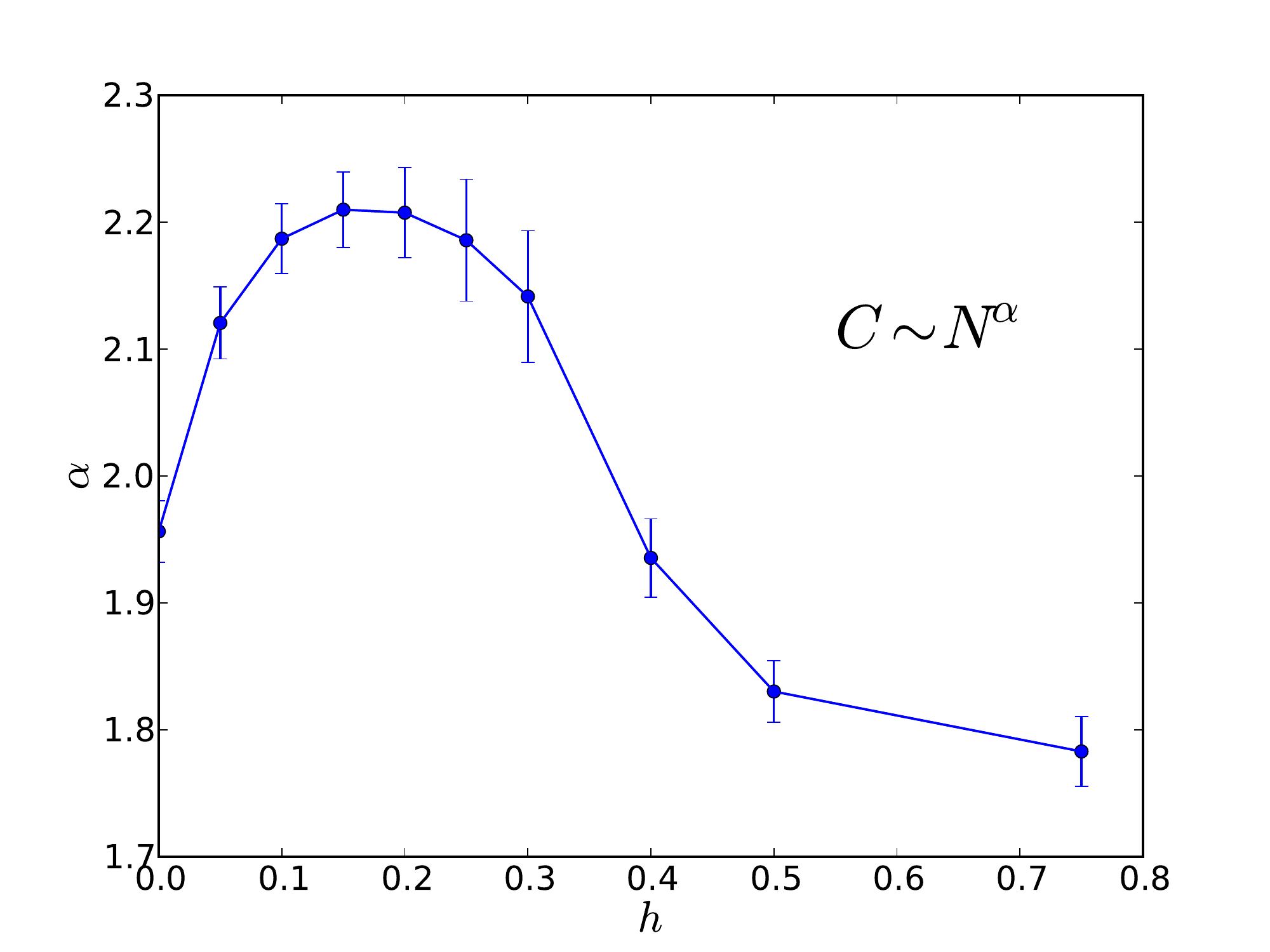}
  \caption{Scaling exponent for the runtime shown in
    Fig. \ref{fig:complexity_L_runtime}, obtained from fitting the
    data to a power law, for different system sizes and various
    fields}
  \label{fig:complexity_scaling_runtime}
\end{figure}

As shown in Figs.~\ref{fig:complexity_L_runtime} and
\ref{fig:complexity_scaling_runtime} the median time to solution
appears to scale polynomially in the number of spins for all values of
the field $h$, including zero field\footnote{The median was calculated
  from $5280$ random instances for each system size and field
  strength. In our implementation we use a library \cite{BDD_library}
  for binary decision diagrams to store these configurations which we
  found to be significantly more efficient than using simple boolean
  tables.}. While faster specialised exact solvers are available
\cite{exact}, this algorithm is not necessarily intended as a general
purpose optimiser, but rather to demonstrate polynomial scaling in the
number of spins at all values of $h$ for typical low-dimensional spin
glass instances.

\section{Hierarchical search}
\label{sec:hierarchical}

\subsection{Motivation}

In this Section we present a general purpose heuristic hierarchical
algorithm for finding the ground state of Ising spin glasses based on
recursively optimising groups of variables. Before describing the
algorithm we motivate why solving groups of variables is significantly
more efficient than solving the whole system at once.

The arguably simplest heuristic algorithm for finding the ground state
is by generating random spin configurations and recording the energy,
in other words random guessing. The probability to find the global
ground state of $N$ spins this way is trivially $2^{-N}$ per guess,
assuming for simplicity a non-degenerate ground state in the
discussion here and below. A more sophisticated way to guess the
solution is to generate random configurations of only $N_r = N - N_g$
spins and for each configuration find the lowest energy of the
remaining $N_g$ spins by some other algorithm, e.g. by enumerating all
possible combinations or any other more optimized algorithm. This
improves the probability of guessing the correct solution, but as the
cost of finding the optimal orientation of the remaining $N_g$
variables may be as much as $2^{N_g}$, we might not have gained
much. This idea can, however, be extended to solving multiple
groups. Let's consider two groups with $N_1$ and $N_2$ spins
respectively, chosen such that spins in one group do not couple to any
of the spins in the other group.  For each random guess of the
remaining $N_r = N - N_1 - N_2$ spins, the complexity of finding the
optimal configuration of both of them with respect to the rest of the
system is $2^{N_1} + 2^{N_2}$, thus reducing the total complexity by
an exponential amount from $2^N=2^{N_r + N_1 + N_2}$ to
$2^{N_r}\left(2^{N_1} + 2^{N_2}\right)$. In our algorithm, described
below, we find a significant reduction in complexity even if spins in
subgroups are coupled and overlap with each other.

\subsection{Optimization of groups}

The above argument provides a basis for a simple algorithm to find the
global ground state by iteratively optimising groups of spins. We
start with a random global state, sequentially pick $M$ groups with
$N_g$ spins each and optimise their configurations by calling some --
as yet unspecified -- solver as follows
\begin{algorithmic}

  \Procedure{Solve}{$ $}

  \State $\text{initialise random spin configuration}$

  \For{$j \in \{1\ldots M\}$}

  \State pick a random spin $i$
  \State $\text{build group $G$ of size $N_g$ around spin $i$}$
  \State $\vec\sigma \gets $ \Call{Solve group}{$G $}
  \State \Call{Update configuration}{$G,\vec \sigma$}

  \EndFor

  \EndProcedure

\end{algorithmic}

Here, \Call{Solve group}{$G $} is a solver that solves the group $G$
(taking into account the interaction with spins outside $G$ to produce
an effective field) and returns an optimized configuration
$\vec\sigma$ for the spins in that group.  The procedure \Call{Update
  configuration}{$G,\vec \sigma$} updates the spins inside $G$ to
configuration $\vec \sigma$; if the solver \Call{Solve group}{} is a
heuristic solver, then \Call{Update configuration}{$G,\vec \sigma$}
only makes this change if the energy is lowered.  Alternatively one
may also consider an algorithm which replaces the group configuration
probabilistically with a Metropolis-type criterion or similar.

If we pick trivial groups of size $N_g = 1$, consisting of a single
spin, the group solver just returns the spin direction which minimises
its energy with respect to its neighbours. For larger groups -- as
will usually be the case -- we can use any arbitrary exact or
heuristic solver, including potentially special purpose classical or
quantum hardware.  We note in passing that in the case of $N_g = 1$,
if the new configuration is accepted probabilistically depending on its
energy this algorithm reduces to simulated annealing.
 
\subsection{Hierarchical recursive algorithm}

If solving a given system in groups is more efficient than solving the
whole system at once, performance can be increased even further by
solving each group by subdividing it recursively into sub-groups, thus
giving a hierarchical version of the algorithm. That is, in the pseudo
code written above, we could use the function $\Call{Solve}$,
restricted to the spins in a group, as the solver $\Call{Solve
  Group}$.  The recursion terminates at some (small) group size, which
is solved by another algorithm.

Note that the hierarchical scheme randomises the configuration of each
group before solving it by optimising subgroups, thus implementing
random local restarts without affecting the global spin
configuration. This randomisation also implies that it makes no sense
to solve a particular group more than once in a row, but rather a new
group should be chosen after one group has been optimized. It should
be emphasised that random restarting is just one possible way to
initialise the state of a group and the one we used here. Other ways
are possible and could be more efficient.

The total complexity of the hierarchical algorithm is dominated by the
number of calls to the solver for the bottom level group rather than
by the group size at each level. This is because for a given group of
size $N_g$, the effort to calculate the local energy and randomise
spins is at most $\mathcal{O}(N_g^2)$ for dense graphs, which is
typically negligible relative to the effort of finding a lower energy
configuration of that group.

\subsection{Selecting groups}

\begin{figure}
  \vspace{0.8cm}
  \includegraphics[width=0.7\columnwidth]{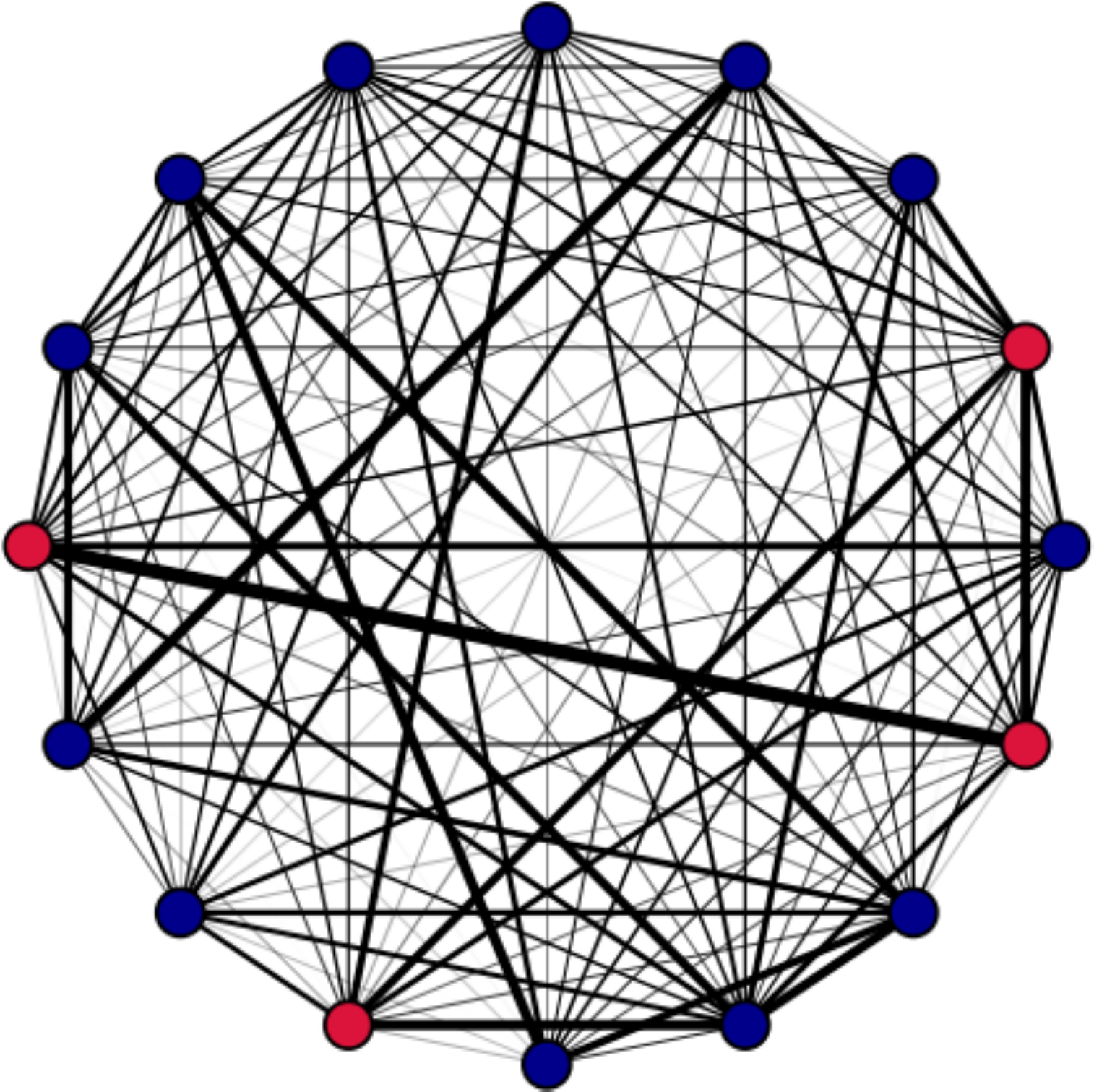}
  \caption{Optimal group, marked in red, of $4$ spins in complete
    graph of $16$ spins with Gaussian disorder.}
  \label{fig:group}
\end{figure}

Up to now, we ignored the hard problem of how to best pick
groups. Here we provide a simple strategy that turned out to work
well. Intuitively, in a well chosen group spins are strongly coupled
to each other and more weakly coupled to the rest of the system, see
Fig.~\ref{fig:group}. We thus build a group $G$ by starting from one
spin and greedily adding spins until the group $G$ has reached the
desired size. We add the spin $i$ that maximises $W_i=\sum_{j\in
  G}|J_{ij}| - \sum_{j\not\in G}|J_{ij}|$, if this maximum is positive
and a random neighbour of one of the spins in $G$ otherwise.

Other ways of building a group may be more effective. For example,
single spins could be added probabilistically, or instead of single
spins we could consider sets of spins which can be added to the
group. Such improvements will be discussed in follow-up work.

\subsection{Results}

To test the performance of our algorithm we compare it to simulated
annealing, which is currently one of the most versatile and efficient
solver for finding ground states of spin glasses. As mentioned above,
simulated annealing is a special case of our algorithm. For our
benchmarks we perform hierarchical search with two levels, using
simulated annealing to solve groups of size $N_1$ with the optimised
configuration being accepted if its energy is lower or the same as the
current configuration.

As a measure of complexity we use the median total number of spin
updates required to find the ground state with a target probability
$p_0 = 0.99$. Since a heuristic algorithm will find the ground state
with some probability $p_s<1$ we may have to repeat the optimization
multiple times if $p_s<p_0$.  Assuming independent repetitions, the
required number of repetitions is $R = \lceil\log{(1 - p_0)}/\log{(1 -
  p_s)}\rceil$. For each set of parameters the probability $p_s$ was
estimated by performing $1024$ repetitions from random initial states.

\begin{table}
\begin{tabular}{|l|lll|l|}
  \hline
  \hline
  $N$ & $M$ & $N_g$ & $S_g$ & $S$\\
  \hline
  32 & 78 & 9 & 1 & 8\\
  72 & 80 & 37 & 5 & 24\\
  128 & 100 & 60 & 7 & 64\\
  200 & 349 & 41 & 4 & 192\\
  288 & 408 & 68 & 6 & 400\\
  392 & 500 & 105 & 13 & 1024\\
  512 & 642 & 129 & 14 & 2048\\
  \hline
  \hline
\end{tabular}
\caption{Optimal parameters for chimera graphs with random bimodal
  disorder. $N$ is the system size, $M$ is the number of groups,
  $N_g$ is the group size, $S_g$ is the number of simulated annealing
  sweeps per group, $S$ is the number of sweeps for plain simulated
  annealing.}
\label{table:chimera_bimodal}
\end{table}

\begin{table}
\begin{tabular}{|l|lll|l|}
  \hline
  \hline
  $N$ & $M$ & $N_g$ & $S_g$ & $S$\\
  \hline
  32 & 28 & 8 & 1 & 4\\
  72 & 237 & 9 & 1 & 4\\
  128 & 388 & 9 & 1 & 4\\
  200 & 197 & 26 & 3 & 1281\\
  288 & 454 & 29 & 3 & 4800\\
  392 & 470 & 27 & 3 & 24576\\
  512 & 718 & 28 & 3 & 131072\\
  \hline
  \hline
\end{tabular}
\caption{Optimal parameters for chimera graphs with cluster bimodal
  disorder. The parameters have the same meaning as in
  Tab. \ref{table:chimera_bimodal}.  }
\label{table:chimera_cluster}
\end{table}

\begin{table}
\begin{tabular}{|l|lll|l|}
  \hline
  \hline
  $N$ & $M$ & $N_g$ & $S_g$ & $S$\\
  \hline
  16 & 72 & 8 & 1 & 4\\
  64 & 314 & 9 & 1 & 48\\
  144 & 273 & 33 & 21 & 891\\
  256 & 573 & 47 & 43 & 30189\\
  \hline
  \hline
\end{tabular}
\caption{Optimal parameters for two dimensional lattices with Gaussian
  disorder. The parameters have the same meaning as in
  Tab. \ref{table:chimera_bimodal}.}
\label{table:2d_Gaussian}
\end{table}

\begin{table}
\begin{tabular}{|l|lll|l|}
  \hline
  \hline
  $N$ & $M$ & $N_g$ & $S_g$ & $S$\\
  \hline
  27 & 39 & 10 & 1 & 5\\
  64 & 118 & 22 & 4 & 45\\
  125 & 232 & 30 & 5 & 512\\
  216 & 271 & 58 & 23 & 2700\\
  343 & 562 & 86 & 42 & 13056\\
  512 & 614 & 113 & 101 & 61440\\
  \hline
  \hline
\end{tabular}
\caption{Optimal parameters for three dimensional lattices with
  Gaussian disorder. The parameters have the same meaning as in
  Tab. \ref{table:chimera_bimodal}.}
\label{table:3d_Gaussian}
\end{table}

\begin{table}
\begin{tabular}{|l|lll|l|}
  \hline
  \hline
  $N$ & $M$ & $N_g$ & $S_g$ & $S$\\
  \hline
  256 & 235 & 56 & 10 & 64\\
  400 & 224 & 119 & 23 & 227\\
  576 & 384 & 103 & 17 & 768\\
  784 & 686 & 208 & 31 & 3506\\
  1024 & 656 & 151 & 29 & 7680\\
  \hline
  \hline
\end{tabular}
\caption{Optimal parameters for two dimensional lattices with bimodal
  disorder. The parameters have the same meaning as in
  Tab. \ref{table:chimera_bimodal}.}
\label{table:2d_bimodal}
\end{table}

For both algorithms and each class and size of problems we optimise
the simulation parameters to minimise the median effort in terms of
single spin updates. For simulated annealing the total effort for a
single repetition is $S N$, there $S$ is the number of sweeps and $N$
is the system size. We used a linear schedule in $\beta = 1/T$ where
the initial and final values of inverse temperature, $\beta_0$ and
$\beta_1$ respectively, as well as the number of sweeps $S$ are chosen
to minimise the total effort. We list the parameters used in Tables
\ref{table:chimera_bimodal} -- \ref{table:2d_bimodal}.

For the hierarchical approach a single repetitions requires a total
effort $M S_g N_g$, where $M$ is the number of groups, $S_g$ is the
number of simulated annealing sweeps per group and $N_g$ is the group
size. The same annealing schedule is used for each group. The values
of $M$, $S_g$ and $N_g$ are chosen to minimise the total effort and
are listed in Tables \ref{table:chimera_bimodal} --
\ref{table:2d_bimodal}

\begin{figure}
  \centering
  \includegraphics[width=0.8\columnwidth]{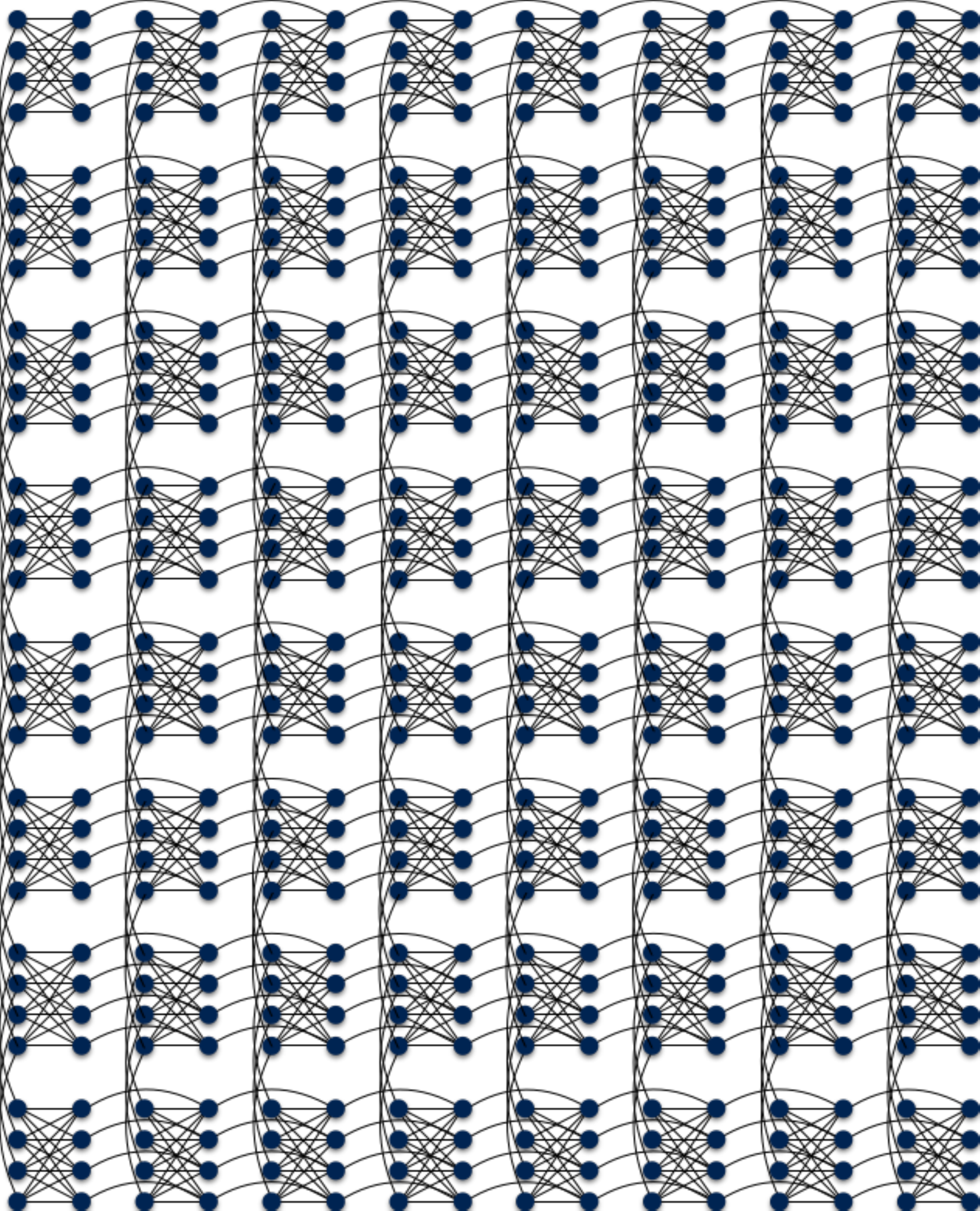}
  \caption{Chimera graph with $512$ spins composed of an $8\times 8$
    grid of unit cell. Each unit cells is a complete bipartite graph
    with $8$ spins.}
  \label{fig:chimera}
\end{figure}

As benchmark problems we used typical spin glass problems on two and
three-dimensional lattices: two-dimensional square lattices, three
dimensional simple cubic lattices\footnote{We used the spin glass
  server \cite{SGS} to computed the exact ground states for three
  dimensional lattices}, and so-called two-dimensional chimera
graphs. The unit cell of the chimera graph \cite{chimera}, shown in
Fig. \ref{fig:chimera}, is a complete bipartite graph with eight
vertices and is coupled to the neighbouring unit cells with four edges
each. Hence, each vertex has either five or six edges corresponding to
four edges to spins within the unit cell and one or two edges to
neighbouring unit cells depending on if it is on the edges of the
graph or in the interior respectively.

One choice of benchmark problems are spin glasses with bimodal
disorder {\it i.e.}, couplings $J_{ij} = \pm 1$ and another choice
will be Gaussian disorder with couplings drawn from a normal
distribution with zero mean and unit variance. In all benchmarks we
choose zero local fields $h=0$.

A special benchmark problem is chimera graphs with cluster structure,
which has recently been proposed as a class of problems to explore an
advantage of quantum annealing over simulated annealing
\cite{chimera_cluster}. In these problems the spins within each unit
cell are coupled ferromagnetically with $J_{ij} = -1$. Of the four
edges connecting neighbouring pairs of unit cells one randomly chosen
edge is assigned a random coupling $J_{ij} = \pm 1$ and the rest is
set to zero.


\begin{figure}
  \centering
  \includegraphics[width=1.0\columnwidth]{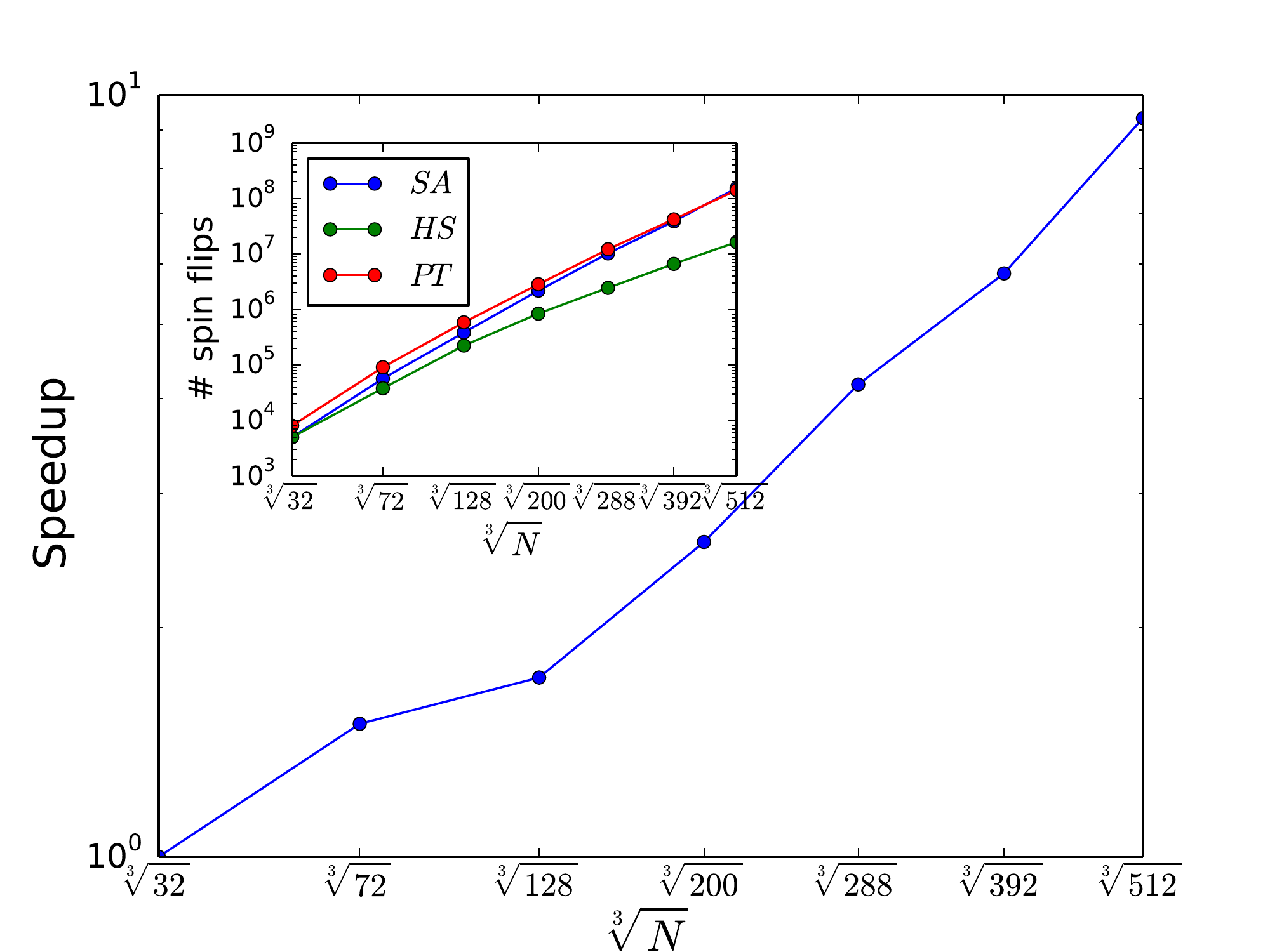}
  \caption{Speedup (ratio of the median number of spin updates) of
    hierarchical search relative to simulated annealing for chimera
    graphs with random bimodal disorder. The inset shows the total
    number of spin updates for simulated annealing (SA), parallel
    tempering (PT) and hierarchical search (HS). For SA, PT and each
    group of HS, $\beta_0 = 0.1$, $\beta_1 = 3$.  Optimal parameters
    for SA and HS are listed in Tab. \ref{table:chimera_bimodal}.}
\label{fig:SA_SAP_speedup_random}
\end{figure}

\begin{figure}
  \centering
  \includegraphics[width=1.0\columnwidth]{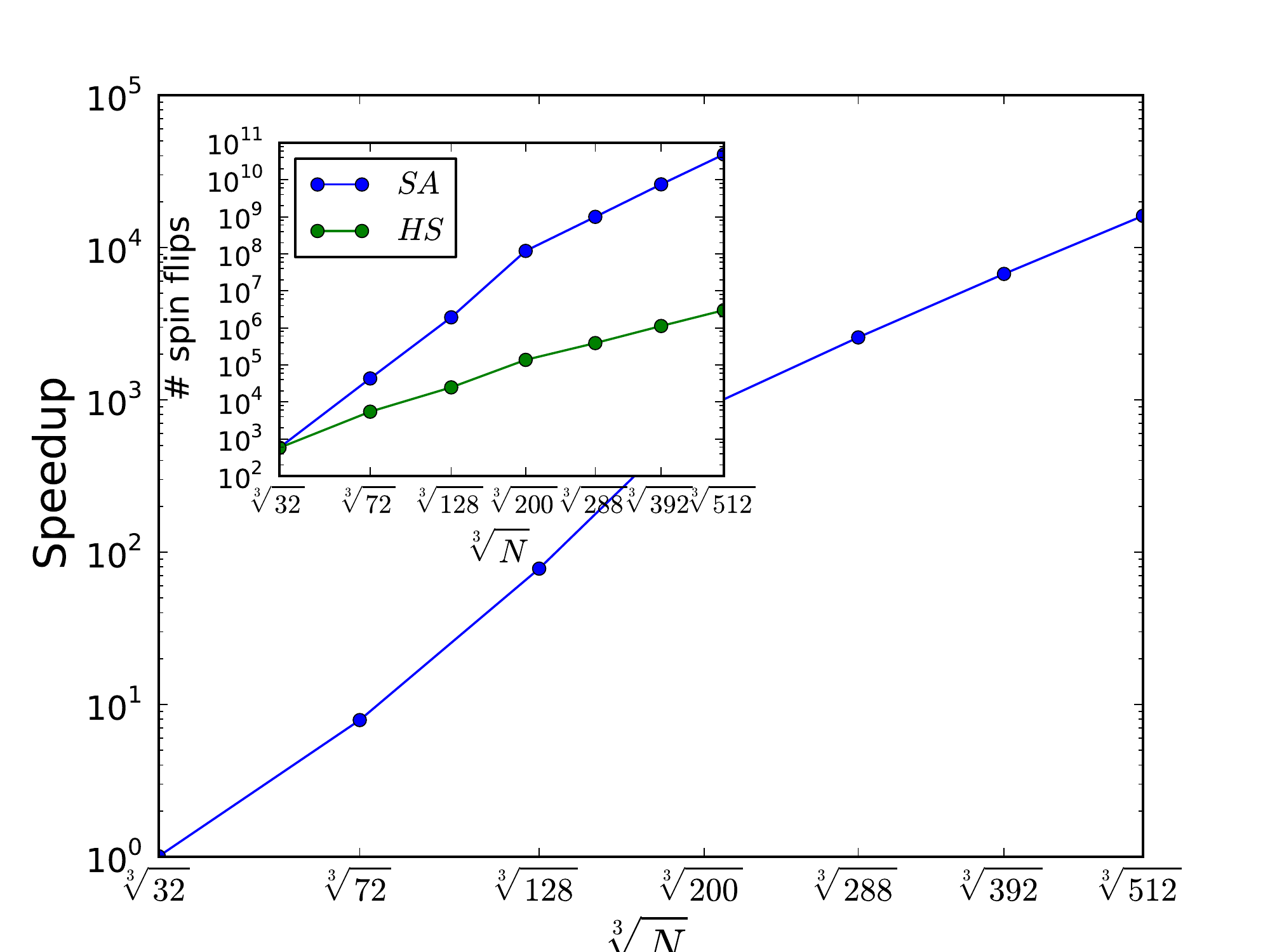}
  \caption{Chimera graphs with cluster bimodal disorder. Speedup and
    the inset are defined the same as in
    Fig. \ref{fig:SA_SAP_speedup_random}. For both plain simulated
    annealing and for each group, $\beta_0 = 0.1$, $\beta_1 =
    3$. Optimal parameters for both algorithms are listed in
    Tab. \ref{table:chimera_cluster}.}
\label{fig:SA_SAP_speedup_cluster}
\end{figure}

\begin{figure}
  \centering
  \includegraphics[width=1.0\columnwidth]{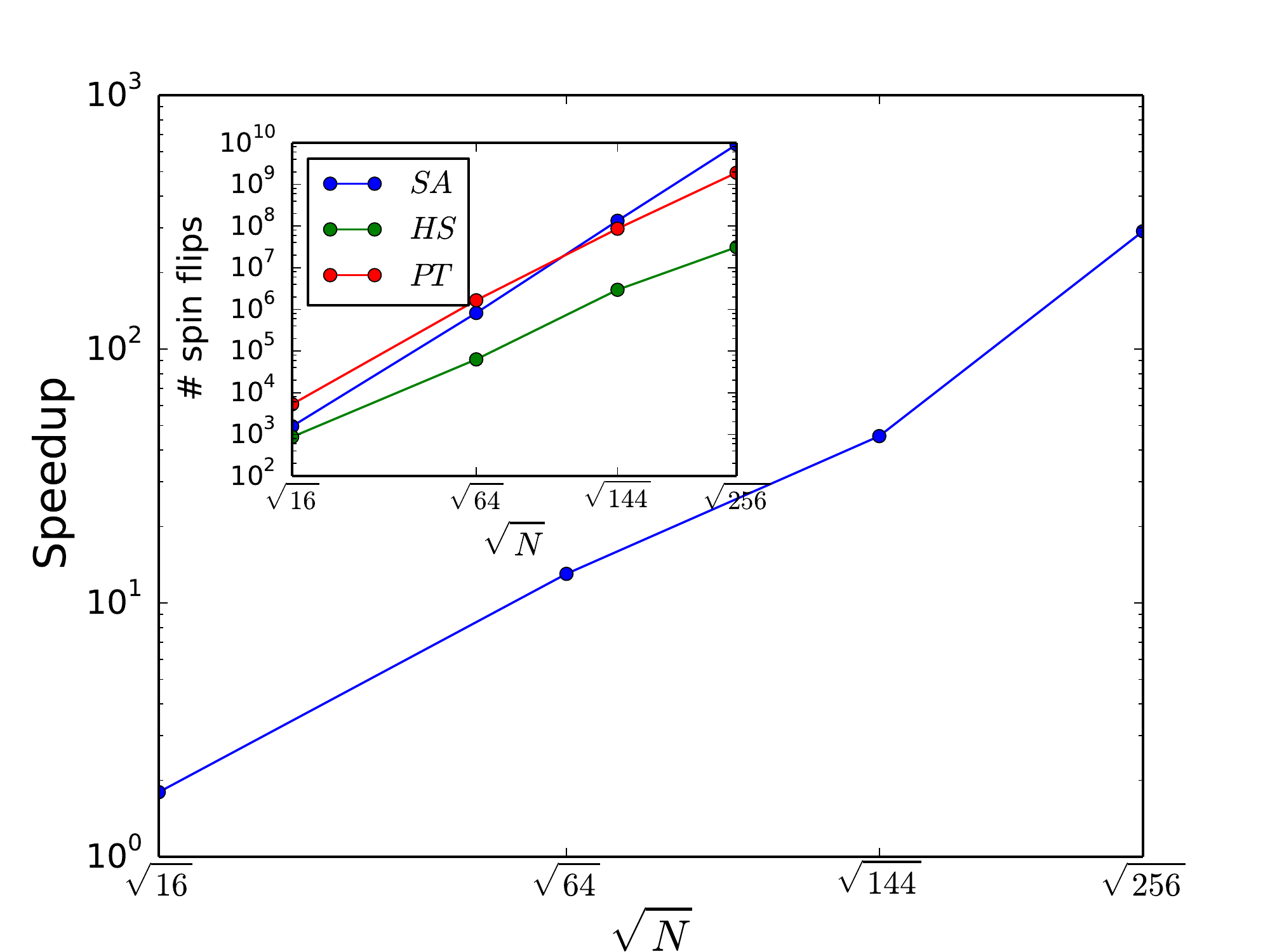}
  \caption{Two dimensional square lattices with Gaussian
    disorder. Speedup and the inset are defined the same as in
    Fig. \ref{fig:SA_SAP_speedup_random}. As the energy gap between
    the ground state and first excited state decreases linearly with
    system size, the final temperature is also reduced with the number
    of spins. $\beta_0 = 0.014$, $\beta_1 = 0.037 N + 2.5$ for plain
    simulated annealing (SA) and parallel tempering (PT) and $\beta_1
    = 0.037 N_g + 2.5$ for each group of the hierarchical algorithm
    (HS). Optimal parameters for SA and HS are listed in
    Tab. \ref{table:2d_Gaussian}.}
\label{fig:SA_SAP_speedup_2DGaussian}
\end{figure}

\begin{figure}
  \centering
  \includegraphics[width=1.0\columnwidth]{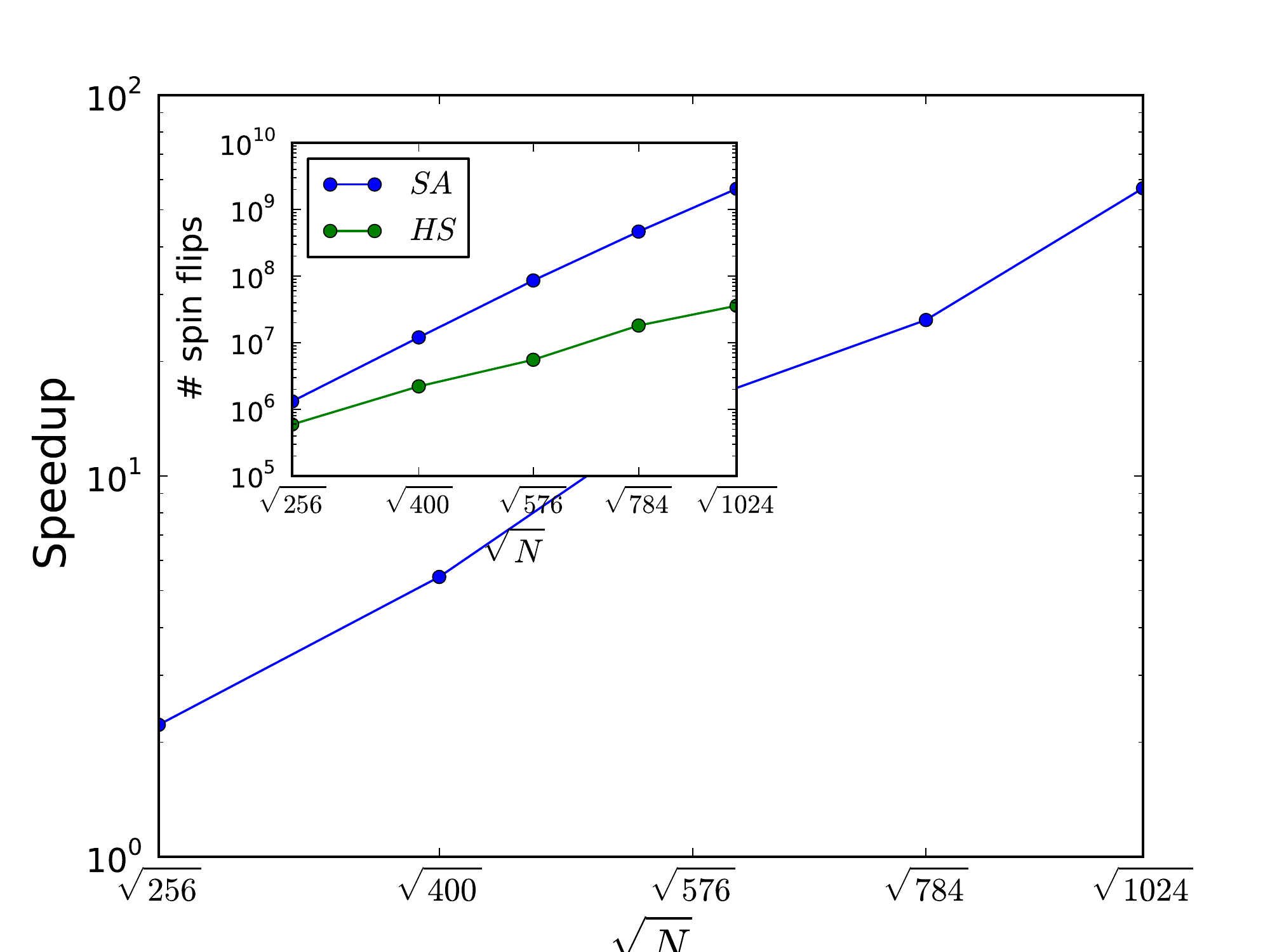}
  \caption{Two dimensional square lattices with bimodal
    disorder. Speedup and the inset are defined the same as in
    Fig. \ref{fig:SA_SAP_speedup_random}. For both plain simulated
    annealing and for each group, $\beta_0 = 0.2$, $\beta_1 =
    3$. Optimal parameters for both algorithms are listed in
    Tab. \ref{table:2d_bimodal}.}
\label{fig:SA_SAP_speedup_2Dbimodal}
\end{figure}

\begin{figure}
  \centering
  \includegraphics[width=1.0\columnwidth]{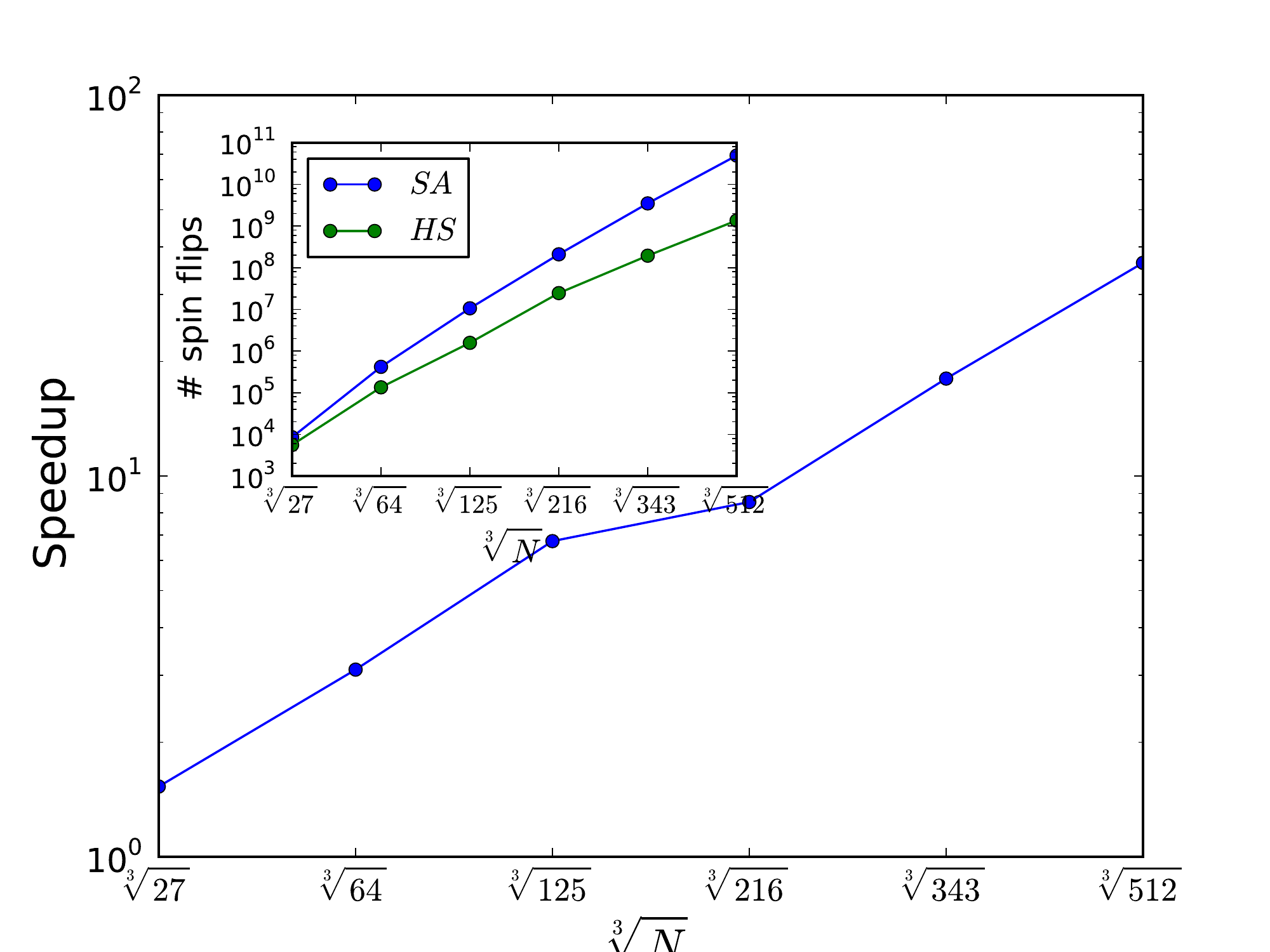}
  \caption{Three dimensional cubic lattices with Gaussian
    disorder. Speedup and the inset are defined the same as in
    Fig. \ref{fig:SA_SAP_speedup_random}. $\beta_0 = 0.05$, $\beta_1 =
    0.028 N + 5.68$ for plain simulated annealing and $\beta_1 = 0.028
    N_g + 5.68$ for each group of the hierarchical algorithm. Optimal
    parameters for both algorithms are listed in
    Tab. \ref{table:3d_Gaussian}.}
\label{fig:SA_SAP_speedup_3DGaussian}
\end{figure}

\begin{figure}
  \centering
  \includegraphics[width=1.0\columnwidth]{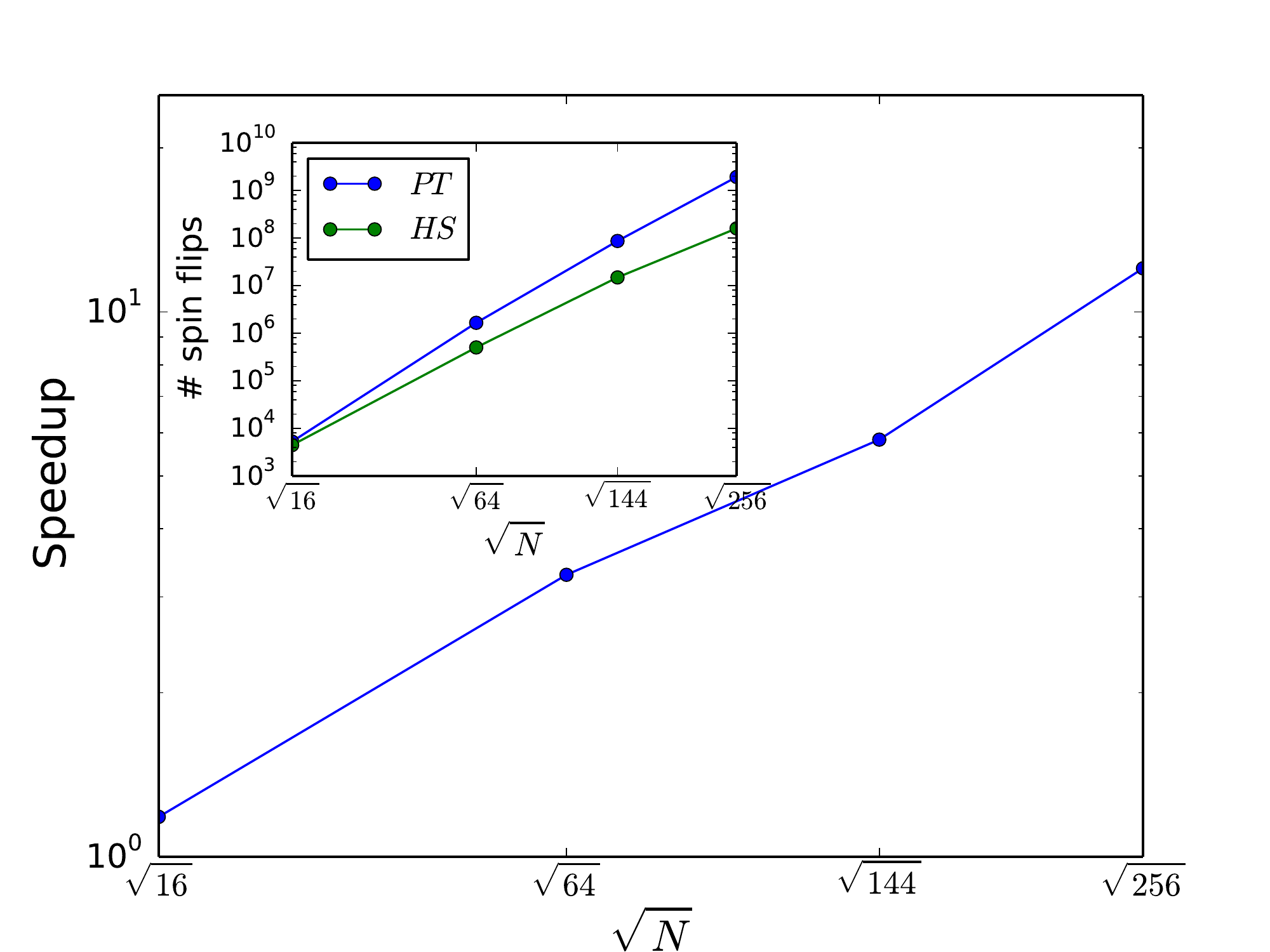}
  \caption{Speedup (ratio of the median number of spin updates) of
    hierarchical search (HS) with parallel tempering as a group solver
    to plain parallel tempering (PT) for two dimensional square
    lattices with Gaussian disorder. $\beta_0 = 0.014$, $\beta_1 =
    0.037 N + 2.5$ for plain parallel tempering and $\beta_1 = 0.037
    N_g + 2.5$ for each group of the hierarchical algorithm.}
\label{fig:PT_PTP_speedup_2DGaussian}
\end{figure}

In all benchmarks we find that hierarchical search performs
significantly better than simulated annealing. The gain is evidently
more significant for problems that are harder for simulated annealing,
such as cluster chimera graphs and systems with Gaussian disorder, see
Fig.~\ref{fig:SA_SAP_speedup_cluster},
\ref{fig:SA_SAP_speedup_2DGaussian} and
\ref{fig:SA_SAP_speedup_3DGaussian} respectively. Random bimodal
disorder is significantly easier for simulated annealing and hence the
speedup on those problems is smaller, although still substantial, see
Fig.~\ref{fig:SA_SAP_speedup_random} and
\ref{fig:SA_SAP_speedup_2Dbimodal} respectively.

As a further comparison was also made with parallel tempering
\cite{PT}, another state the art method for finding ground states of
Ising spin glasses. For each class and problem size, the total number
of replicas and sweeps per replica was optimised minimising the median
total number of spin updates. For a single repetition the total effort
is $N_R S N$, where $N_R$ is the number of replicas, $S$ is the number
of sweeps per replica and $N$ is the system size. On chimera graphs
its performance is very similar to simulated annealing, see
Fig.~\ref{fig:SA_SAP_speedup_random}. On two dimensional lattices with
Gaussian disorder it performs slightly better, see
Fig.~\ref{fig:SA_SAP_speedup_2DGaussian}. However, analogous to
simulated annealing its performance can be significantly improved by
optimising groups of spins rather than the whole system at once, see
Fig.~\ref{fig:PT_PTP_speedup_2DGaussian}. Note that although in all
cases the advantage of hierarchical search over plain simulated
annealing and parallel tempering grows with problem size, a spin
update is effectively more costly due to the additional overhead of
randomising the spins and computing the energy of a group. However,
the difference is typically insignificant. For example, the wall clock
time per spin update is only about 7\% higher than plain simulated
annealing for $8\times8\times8$ 3D lattices with Gaussian disorder ran
with optimal parameters.

\section{Conclusion}

It has long been established that the complexity of finding grounds
states of spin glasses is strongly dependent on the ensemble of
couplings and is in the worst case NP-hard. However, whilst the the
most trivial cases, like the ferromagnetic Ising model, are relatively
evident, the hardest problems are far more
elusive \cite{2014Katzgraber}.

One way to look for hard problems is by sampling randomly distributed
couplings. Although this approach certainly includes such problems, in
this work we presented strong numerical evidence that the average
complexity of low-dimensional spin glasses with randomly distributed
couplings is actually polynomial in the number of spins and looking
for hard problems in such a large ensemble might be next to
futile. Another way to generate hard cases is to map 3-SAT problems at
the critical clause to variable ratio \cite{Mitchell1992}, where
previous studies have shown evidence of a universal peak in
complexity, to the Ising model. Further studies are to be done in this
direction.

Our most significant result reported here is a hierarchical approach
as a way to potentially improve the performance of a given algorithm
for finding ground states of Ising spin glasses. With simulated
annealing as a reference solver, on all our benchmark instances we
find that optimising groups of spins is significantly more efficient
than solving the whole system at once.

It should be noted that approaches other than simulated annealing can
be used at the bottom level of the hierarchical solver. Suppose for
some class of problems, another algorithm (or special purpose
classical or quantum device) outperforms simulated annealing. In that
case, we can use that algorithm or device at the lowest level. Let
$T_0$ denote the time annealing takes to optimise the bottom level. If
it is now replaced by a device which takes time, including
communication overhead, $T_1 \ll T_0$, we expect the potential speedup
to the whole algorithm to be $\sim T_0/T_1$. As the complexity of
finding the ground state scales exponentially with the number of
spins, this can be significant even for small groups.

Although we limited our investigation to spin glasses, similar ideas
can be applied directly to other problems such as machine learning,
protein folding, travelling salesman etc. by constraining groups of
variables independently relative to the rest of the system. We will
address such applications in a follow-up work.

\section{Acknowledgements}

This work was supported by Microsoft Research, the Swiss National
Competence Center in Research NCCR QSIT and the ERC Advanced Grant
SIMCOFE. We thank Bruichladdich Dist. for inspiration and M. Freedman,
L. Gamper, I. Hen, J. Imriska, S. Isakov, H.G. Katzgraber,
A. Kosenkov, J. Osorio, I. Pizorn, D. Poulin, T. R{\o}nnow, A.
Soluyanov, and D. Steiger for fruitful discussions. Simulations were
performed on the Monch and Brutus clusters of ETH Zurich.

\bibliography{opt}

\begin{thebibliography}{35}%
\makeatletter
\providecommand \@ifxundefined [1]{%
 \@ifx{#1\undefined}
}%
\providecommand \@ifnum [1]{%
 \ifnum #1\expandafter \@firstoftwo
 \else \expandafter \@secondoftwo
 \fi
}%
\providecommand \@ifx [1]{%
 \ifx #1\expandafter \@firstoftwo
 \else \expandafter \@secondoftwo
 \fi
}%
\providecommand \natexlab [1]{#1}%
\providecommand \enquote  [1]{``#1''}%
\providecommand \bibnamefont  [1]{#1}%
\providecommand \bibfnamefont [1]{#1}%
\providecommand \citenamefont [1]{#1}%
\providecommand \href@noop [0]{\@secondoftwo}%
\providecommand \href [0]{\begingroup \@sanitize@url \@href}%
\providecommand \@href[1]{\@@startlink{#1}\@@href}%
\providecommand \@@href[1]{\endgroup#1\@@endlink}%
\providecommand \@sanitize@url [0]{\catcode `\\12\catcode `\$12\catcode
  `\&12\catcode `\#12\catcode `\^12\catcode `\_12\catcode `\%12\relax}%
\providecommand \@@startlink[1]{}%
\providecommand \@@endlink[0]{}%
\providecommand \url  [0]{\begingroup\@sanitize@url \@url }%
\providecommand \@url [1]{\endgroup\@href {#1}{\urlprefix }}%
\providecommand \urlprefix  [0]{URL }%
\providecommand \Eprint [0]{\href }%
\providecommand \doibase [0]{http://dx.doi.org/}%
\providecommand \selectlanguage [0]{\@gobble}%
\providecommand \bibinfo  [0]{\@secondoftwo}%
\providecommand \bibfield  [0]{\@secondoftwo}%
\providecommand \translation [1]{[#1]}%
\providecommand \BibitemOpen [0]{}%
\providecommand \bibitemStop [0]{}%
\providecommand \bibitemNoStop [0]{.\EOS\space}%
\providecommand \EOS [0]{\spacefactor3000\relax}%
\providecommand \BibitemShut  [1]{\csname bibitem#1\endcsname}%
\let\auto@bib@innerbib\@empty
\bibitem [{\citenamefont {Binder}\ and\ \citenamefont
  {Young}(1986)}]{BinderYoung}%
  \BibitemOpen
  \bibfield  {author} {\bibinfo {author} {\bibfnamefont {K.}~\bibnamefont
  {Binder}}\ and\ \bibinfo {author} {\bibfnamefont {A.~P.}\ \bibnamefont
  {Young}},\ }\href {\doibase 10.1103/RevModPhys.58.801} {\bibfield  {journal}
  {\bibinfo  {journal} {Rev. Mod. Phys.}\ }\textbf {\bibinfo {volume} {58}},\
  \bibinfo {pages} {801} (\bibinfo {year} {1986})}\BibitemShut {NoStop}%
\bibitem [{\citenamefont {Barahona}(1982)}]{Barahona1982}%
  \BibitemOpen
  \bibfield  {author} {\bibinfo {author} {\bibfnamefont {F.}~\bibnamefont
  {Barahona}},\ }\href {http://stacks.iop.org/0305-4470/15/i=10/a=028}
  {\bibfield  {journal} {\bibinfo  {journal} {Journal of Physics A:
  Mathematical and General}\ }\textbf {\bibinfo {volume} {15}},\ \bibinfo
  {pages} {3241} (\bibinfo {year} {1982})}\BibitemShut {NoStop}%
\bibitem [{\citenamefont {Lucas}(2014)}]{Lucas2014}%
  \BibitemOpen
  \bibfield  {author} {\bibinfo {author} {\bibfnamefont {A.}~\bibnamefont
  {Lucas}},\ }\href {\doibase 10.3389/fphy.2014.00005} {\bibfield  {journal}
  {\bibinfo  {journal} {Frontiers in Physics}\ }\textbf {\bibinfo {volume} {2}}
  (\bibinfo {year} {2014}),\ 10.3389/fphy.2014.00005}\BibitemShut {NoStop}%
\bibitem [{\citenamefont {Harris}\ \emph {et~al.}(2010)\citenamefont {Harris},
  \citenamefont {Johnson}, \citenamefont {Lanting}, \citenamefont {Berkley},
  \citenamefont {Johansson}, \citenamefont {Bunyk}, \citenamefont {Tolkacheva},
  \citenamefont {Ladizinsky}, \citenamefont {Ladizinsky}, \citenamefont {Oh},
  \citenamefont {Cioata}, \citenamefont {Perminov}, \citenamefont {Spear},
  \citenamefont {Enderud}, \citenamefont {Rich}, \citenamefont {Uchaikin},
  \citenamefont {Thom}, \citenamefont {Chapple}, \citenamefont {Wang},
  \citenamefont {Wilson}, \citenamefont {Amin}, \citenamefont {Dickson},
  \citenamefont {Karimi}, \citenamefont {Macready}, \citenamefont {Truncik},\
  and\ \citenamefont {Rose}}]{Harris2010}%
  \BibitemOpen
  \bibfield  {author} {\bibinfo {author} {\bibfnamefont {R.}~\bibnamefont
  {Harris}}, \bibinfo {author} {\bibfnamefont {M.~W.}\ \bibnamefont {Johnson}},
  \bibinfo {author} {\bibfnamefont {T.}~\bibnamefont {Lanting}}, \bibinfo
  {author} {\bibfnamefont {A.~J.}\ \bibnamefont {Berkley}}, \bibinfo {author}
  {\bibfnamefont {J.}~\bibnamefont {Johansson}}, \bibinfo {author}
  {\bibfnamefont {P.}~\bibnamefont {Bunyk}}, \bibinfo {author} {\bibfnamefont
  {E.}~\bibnamefont {Tolkacheva}}, \bibinfo {author} {\bibfnamefont
  {E.}~\bibnamefont {Ladizinsky}}, \bibinfo {author} {\bibfnamefont
  {N.}~\bibnamefont {Ladizinsky}}, \bibinfo {author} {\bibfnamefont
  {T.}~\bibnamefont {Oh}}, \bibinfo {author} {\bibfnamefont {F.}~\bibnamefont
  {Cioata}}, \bibinfo {author} {\bibfnamefont {I.}~\bibnamefont {Perminov}},
  \bibinfo {author} {\bibfnamefont {P.}~\bibnamefont {Spear}}, \bibinfo
  {author} {\bibfnamefont {C.}~\bibnamefont {Enderud}}, \bibinfo {author}
  {\bibfnamefont {C.}~\bibnamefont {Rich}}, \bibinfo {author} {\bibfnamefont
  {S.}~\bibnamefont {Uchaikin}}, \bibinfo {author} {\bibfnamefont {M.~C.}\
  \bibnamefont {Thom}}, \bibinfo {author} {\bibfnamefont {E.~M.}\ \bibnamefont
  {Chapple}}, \bibinfo {author} {\bibfnamefont {J.}~\bibnamefont {Wang}},
  \bibinfo {author} {\bibfnamefont {B.}~\bibnamefont {Wilson}}, \bibinfo
  {author} {\bibfnamefont {M.~H.~S.}\ \bibnamefont {Amin}}, \bibinfo {author}
  {\bibfnamefont {N.}~\bibnamefont {Dickson}}, \bibinfo {author} {\bibfnamefont
  {K.}~\bibnamefont {Karimi}}, \bibinfo {author} {\bibfnamefont
  {B.}~\bibnamefont {Macready}}, \bibinfo {author} {\bibfnamefont {C.~J.~S.}\
  \bibnamefont {Truncik}}, \ and\ \bibinfo {author} {\bibfnamefont
  {G.}~\bibnamefont {Rose}},\ }\href {\doibase 10.1103/PhysRevB.82.024511}
  {\bibfield  {journal} {\bibinfo  {journal} {Phys. Rev. B}\ }\textbf {\bibinfo
  {volume} {82}},\ \bibinfo {pages} {024511} (\bibinfo {year}
  {2010})}\BibitemShut {NoStop}%
\bibitem [{\citenamefont {Johnson}\ \emph {et~al.}(2010)\citenamefont
  {Johnson}, \citenamefont {Bunyk}, \citenamefont {Maibaum}, \citenamefont
  {Tolkacheva}, \citenamefont {Berkley}, \citenamefont {Chapple}, \citenamefont
  {Harris}, \citenamefont {Johansson}, \citenamefont {Lanting}, \citenamefont
  {Perminov}, \citenamefont {Ladizinsky}, \citenamefont {Oh},\ and\
  \citenamefont {Rose}}]{0953-2048-23-6-065004}%
  \BibitemOpen
  \bibfield  {author} {\bibinfo {author} {\bibfnamefont {M.~W.}\ \bibnamefont
  {Johnson}}, \bibinfo {author} {\bibfnamefont {P.}~\bibnamefont {Bunyk}},
  \bibinfo {author} {\bibfnamefont {F.}~\bibnamefont {Maibaum}}, \bibinfo
  {author} {\bibfnamefont {E.}~\bibnamefont {Tolkacheva}}, \bibinfo {author}
  {\bibfnamefont {A.~J.}\ \bibnamefont {Berkley}}, \bibinfo {author}
  {\bibfnamefont {E.~M.}\ \bibnamefont {Chapple}}, \bibinfo {author}
  {\bibfnamefont {R.}~\bibnamefont {Harris}}, \bibinfo {author} {\bibfnamefont
  {J.}~\bibnamefont {Johansson}}, \bibinfo {author} {\bibfnamefont
  {T.}~\bibnamefont {Lanting}}, \bibinfo {author} {\bibfnamefont
  {I.}~\bibnamefont {Perminov}}, \bibinfo {author} {\bibfnamefont
  {E.}~\bibnamefont {Ladizinsky}}, \bibinfo {author} {\bibfnamefont
  {T.}~\bibnamefont {Oh}}, \ and\ \bibinfo {author} {\bibfnamefont
  {G.}~\bibnamefont {Rose}},\ }\href
  {http://stacks.iop.org/0953-2048/23/i=6/a=065004} {\bibfield  {journal}
  {\bibinfo  {journal} {Superconductor Science and Technology}\ }\textbf
  {\bibinfo {volume} {23}},\ \bibinfo {pages} {065004} (\bibinfo {year}
  {2010})}\BibitemShut {NoStop}%
\bibitem [{\citenamefont {Berkley}\ \emph {et~al.}(2010)\citenamefont
  {Berkley}, \citenamefont {Johnson}, \citenamefont {Bunyk}, \citenamefont
  {Harris}, \citenamefont {Johansson}, \citenamefont {Lanting}, \citenamefont
  {Ladizinsky}, \citenamefont {Tolkacheva}, \citenamefont {Amin},\ and\
  \citenamefont {Rose}}]{berkley2010scalable}%
  \BibitemOpen
  \bibfield  {author} {\bibinfo {author} {\bibfnamefont {A.~J.}\ \bibnamefont
  {Berkley}}, \bibinfo {author} {\bibfnamefont {M.~W.}\ \bibnamefont
  {Johnson}}, \bibinfo {author} {\bibfnamefont {P.}~\bibnamefont {Bunyk}},
  \bibinfo {author} {\bibfnamefont {R.}~\bibnamefont {Harris}}, \bibinfo
  {author} {\bibfnamefont {J.}~\bibnamefont {Johansson}}, \bibinfo {author}
  {\bibfnamefont {T.}~\bibnamefont {Lanting}}, \bibinfo {author} {\bibfnamefont
  {E.}~\bibnamefont {Ladizinsky}}, \bibinfo {author} {\bibfnamefont
  {E.}~\bibnamefont {Tolkacheva}}, \bibinfo {author} {\bibfnamefont {M.~H.~S.}\
  \bibnamefont {Amin}}, \ and\ \bibinfo {author} {\bibfnamefont
  {G.}~\bibnamefont {Rose}},\ }\href {\doibase 10.1088/0953-2048/23/10/105014}
  {\bibfield  {journal} {\bibinfo  {journal} {Superconductor Science and
  Technology}\ }\textbf {\bibinfo {volume} {23}},\ \bibinfo {pages} {105014}
  (\bibinfo {year} {2010})}\BibitemShut {NoStop}%
\bibitem [{\citenamefont {Johnson}\ \emph {et~al.}(2011)\citenamefont
  {Johnson}, \citenamefont {Amin}, \citenamefont {Gildert}, \citenamefont
  {Lanting}, \citenamefont {Hamze}, \citenamefont {Dickson}, \citenamefont
  {Harris}, \citenamefont {Berkley}, \citenamefont {Johansson}, \citenamefont
  {Bunyk}, \citenamefont {Chapple}, \citenamefont {Enderud}, \citenamefont
  {Hilton}, \citenamefont {Karimi}, \citenamefont {Ladizinsky}, \citenamefont
  {Ladizinsky}, \citenamefont {Oh}, \citenamefont {Perminov}, \citenamefont
  {Rich}, \citenamefont {Thom}, \citenamefont {Tolkacheva}, \citenamefont
  {Truncik}, \citenamefont {Uchaikin}, \citenamefont {Wang}, \citenamefont
  {Wilson},\ and\ \citenamefont {Rose}}]{Johnson2011}%
  \BibitemOpen
  \bibfield  {author} {\bibinfo {author} {\bibfnamefont {M.~W.}\ \bibnamefont
  {Johnson}}, \bibinfo {author} {\bibfnamefont {M.~H.~S.}\ \bibnamefont
  {Amin}}, \bibinfo {author} {\bibfnamefont {S.}~\bibnamefont {Gildert}},
  \bibinfo {author} {\bibfnamefont {T.}~\bibnamefont {Lanting}}, \bibinfo
  {author} {\bibfnamefont {F.}~\bibnamefont {Hamze}}, \bibinfo {author}
  {\bibfnamefont {N.}~\bibnamefont {Dickson}}, \bibinfo {author} {\bibfnamefont
  {R.}~\bibnamefont {Harris}}, \bibinfo {author} {\bibfnamefont {A.~J.}\
  \bibnamefont {Berkley}}, \bibinfo {author} {\bibfnamefont {J.}~\bibnamefont
  {Johansson}}, \bibinfo {author} {\bibfnamefont {P.}~\bibnamefont {Bunyk}},
  \bibinfo {author} {\bibfnamefont {E.~M.}\ \bibnamefont {Chapple}}, \bibinfo
  {author} {\bibfnamefont {C.}~\bibnamefont {Enderud}}, \bibinfo {author}
  {\bibfnamefont {J.~P.}\ \bibnamefont {Hilton}}, \bibinfo {author}
  {\bibfnamefont {K.}~\bibnamefont {Karimi}}, \bibinfo {author} {\bibfnamefont
  {E.}~\bibnamefont {Ladizinsky}}, \bibinfo {author} {\bibfnamefont
  {N.}~\bibnamefont {Ladizinsky}}, \bibinfo {author} {\bibfnamefont
  {T.}~\bibnamefont {Oh}}, \bibinfo {author} {\bibfnamefont {I.}~\bibnamefont
  {Perminov}}, \bibinfo {author} {\bibfnamefont {C.}~\bibnamefont {Rich}},
  \bibinfo {author} {\bibfnamefont {M.~C.}\ \bibnamefont {Thom}}, \bibinfo
  {author} {\bibfnamefont {E.}~\bibnamefont {Tolkacheva}}, \bibinfo {author}
  {\bibfnamefont {C.~J.~S.}\ \bibnamefont {Truncik}}, \bibinfo {author}
  {\bibfnamefont {S.}~\bibnamefont {Uchaikin}}, \bibinfo {author}
  {\bibfnamefont {J.}~\bibnamefont {Wang}}, \bibinfo {author} {\bibfnamefont
  {B.}~\bibnamefont {Wilson}}, \ and\ \bibinfo {author} {\bibfnamefont
  {G.}~\bibnamefont {Rose}},\ }\href {\doibase 10.1038/nature10012} {\bibfield
  {journal} {\bibinfo  {journal} {Nature}\ }\textbf {\bibinfo {volume} {473}},\
  \bibinfo {pages} {194} (\bibinfo {year} {2011})}\BibitemShut {NoStop}%
\bibitem [{\citenamefont {Kadowaki}\ and\ \citenamefont
  {Nishimori}(1998)}]{Kadowaki1998}%
  \BibitemOpen
  \bibfield  {author} {\bibinfo {author} {\bibfnamefont {T.}~\bibnamefont
  {Kadowaki}}\ and\ \bibinfo {author} {\bibfnamefont {H.}~\bibnamefont
  {Nishimori}},\ }\href {\doibase 10.1103/PhysRevE.58.5355} {\bibfield
  {journal} {\bibinfo  {journal} {Phys. Rev. E}\ }\textbf {\bibinfo {volume}
  {58}},\ \bibinfo {pages} {5355} (\bibinfo {year} {1998})}\BibitemShut
  {NoStop}%
\bibitem [{\citenamefont {Shin}\ \emph {et~al.}(2014)\citenamefont {Shin},
  \citenamefont {Smith}, \citenamefont {Smolin},\ and\ \citenamefont
  {Vazirani}}]{SSSV}%
  \BibitemOpen
  \bibfield  {author} {\bibinfo {author} {\bibfnamefont {S.~W.}\ \bibnamefont
  {Shin}}, \bibinfo {author} {\bibfnamefont {G.}~\bibnamefont {Smith}},
  \bibinfo {author} {\bibfnamefont {J.~A.}\ \bibnamefont {Smolin}}, \ and\
  \bibinfo {author} {\bibfnamefont {U.}~\bibnamefont {Vazirani}},\ }\href
  {http://arXiv.org/abs/1401.7087} {\bibfield  {journal} {\bibinfo  {journal}
  {arXiv:1401.7087}\ } (\bibinfo {year} {2014})}\BibitemShut {NoStop}%
\bibitem [{\citenamefont {R{\o}nnow}\ \emph {et~al.}(2014)\citenamefont
  {R{\o}nnow}, \citenamefont {Wang}, \citenamefont {Job}, \citenamefont
  {Boixo}, \citenamefont {Isakov}, \citenamefont {Wecker}, \citenamefont
  {Martinis}, \citenamefont {Lidar},\ and\ \citenamefont
  {Troyer}}]{Ronnow2014}%
  \BibitemOpen
  \bibfield  {author} {\bibinfo {author} {\bibfnamefont {T.~F.}\ \bibnamefont
  {R{\o}nnow}}, \bibinfo {author} {\bibfnamefont {Z.}~\bibnamefont {Wang}},
  \bibinfo {author} {\bibfnamefont {J.}~\bibnamefont {Job}}, \bibinfo {author}
  {\bibfnamefont {S.}~\bibnamefont {Boixo}}, \bibinfo {author} {\bibfnamefont
  {S.~V.}\ \bibnamefont {Isakov}}, \bibinfo {author} {\bibfnamefont
  {D.}~\bibnamefont {Wecker}}, \bibinfo {author} {\bibfnamefont {J.~M.}\
  \bibnamefont {Martinis}}, \bibinfo {author} {\bibfnamefont {D.~A.}\
  \bibnamefont {Lidar}}, \ and\ \bibinfo {author} {\bibfnamefont
  {M.}~\bibnamefont {Troyer}},\ }\href {\doibase 10.1126/science.1252319}
  {\bibfield  {journal} {\bibinfo  {journal} {Science}\ }\textbf {\bibinfo
  {volume} {345}},\ \bibinfo {pages} {420} (\bibinfo {year}
  {2014})}\BibitemShut {NoStop}%
\bibitem [{\citenamefont {Bieche}\ \emph {et~al.}(1980)\citenamefont {Bieche},
  \citenamefont {Uhry}, \citenamefont {Maynard},\ and\ \citenamefont
  {Rammal}}]{matching}%
  \BibitemOpen
  \bibfield  {author} {\bibinfo {author} {\bibfnamefont {L.}~\bibnamefont
  {Bieche}}, \bibinfo {author} {\bibfnamefont {J.~P.}\ \bibnamefont {Uhry}},
  \bibinfo {author} {\bibfnamefont {R.}~\bibnamefont {Maynard}}, \ and\
  \bibinfo {author} {\bibfnamefont {R.}~\bibnamefont {Rammal}},\ }\href
  {http://stacks.iop.org/0305-4470/13/i=8/a=005} {\bibfield  {journal}
  {\bibinfo  {journal} {Journal of Physics A: Mathematical and General}\
  }\textbf {\bibinfo {volume} {13}},\ \bibinfo {pages} {2553} (\bibinfo {year}
  {1980})}\BibitemShut {NoStop}%
\bibitem [{\citenamefont {Hartmann}\ and\ \citenamefont
  {Young}(2001)}]{Hartmann01}%
  \BibitemOpen
  \bibfield  {author} {\bibinfo {author} {\bibfnamefont {A.~K.}\ \bibnamefont
  {Hartmann}}\ and\ \bibinfo {author} {\bibfnamefont {A.~P.}\ \bibnamefont
  {Young}},\ }\href {\doibase 10.1103/PhysRevB.64.180404} {\bibfield  {journal}
  {\bibinfo  {journal} {Phys. Rev. B}\ }\textbf {\bibinfo {volume} {64}},\
  \bibinfo {pages} {180404} (\bibinfo {year} {2001})}\BibitemShut {NoStop}%
\bibitem [{\citenamefont {McMillan}(1984)}]{mcmillan}%
  \BibitemOpen
  \bibfield  {author} {\bibinfo {author} {\bibfnamefont {W.~L.}\ \bibnamefont
  {McMillan}},\ }\href {\doibase 10.1103/PhysRevB.30.476} {\bibfield  {journal}
  {\bibinfo  {journal} {Phys. Rev. B}\ }\textbf {\bibinfo {volume} {30}},\
  \bibinfo {pages} {476} (\bibinfo {year} {1984})}\BibitemShut {NoStop}%
\bibitem [{\citenamefont {Leuzzi}\ \emph {et~al.}(2009)\citenamefont {Leuzzi},
  \citenamefont {Parisi}, \citenamefont {Ricci-Tersenghi},\ and\ \citenamefont
  {Ruiz-Lorenzo}}]{field0}%
  \BibitemOpen
  \bibfield  {author} {\bibinfo {author} {\bibfnamefont {L.}~\bibnamefont
  {Leuzzi}}, \bibinfo {author} {\bibfnamefont {G.}~\bibnamefont {Parisi}},
  \bibinfo {author} {\bibfnamefont {F.}~\bibnamefont {Ricci-Tersenghi}}, \ and\
  \bibinfo {author} {\bibfnamefont {J.~J.}\ \bibnamefont {Ruiz-Lorenzo}},\
  }\href {\doibase 10.1103/PhysRevLett.103.267201} {\bibfield  {journal}
  {\bibinfo  {journal} {Phys. Rev. Lett.}\ }\textbf {\bibinfo {volume} {103}},\
  \bibinfo {pages} {267201} (\bibinfo {year} {2009})}\BibitemShut {NoStop}%
\bibitem [{\citenamefont {Baños}\ \emph {et~al.}(2012)\citenamefont {Baños},
  \citenamefont {Cruz}, \citenamefont {Fernandez}, \citenamefont {Gil-Narvion},
  \citenamefont {Gordillo-Guerrero}, \citenamefont {Guidetti}, \citenamefont
  {Iñiguez}, \citenamefont {Maiorano}, \citenamefont {Marinari}, \citenamefont
  {Martin-Mayor}, \citenamefont {Monforte-Garcia}, \citenamefont
  {Muñoz~Sudupe}, \citenamefont {Navarro}, \citenamefont {Parisi},
  \citenamefont {Perez-Gaviro}, \citenamefont {Ruiz-Lorenzo}, \citenamefont
  {Schifano}, \citenamefont {Seoane}, \citenamefont {Tarancon}, \citenamefont
  {Tellez}, \citenamefont {Tripiccione},\ and\ \citenamefont
  {Yllanes}}]{field1}%
  \BibitemOpen
  \bibfield  {author} {\bibinfo {author} {\bibfnamefont {R.~A.}\ \bibnamefont
  {Baños}}, \bibinfo {author} {\bibfnamefont {A.}~\bibnamefont {Cruz}},
  \bibinfo {author} {\bibfnamefont {L.~A.}\ \bibnamefont {Fernandez}}, \bibinfo
  {author} {\bibfnamefont {J.~M.}\ \bibnamefont {Gil-Narvion}}, \bibinfo
  {author} {\bibfnamefont {A.}~\bibnamefont {Gordillo-Guerrero}}, \bibinfo
  {author} {\bibfnamefont {M.}~\bibnamefont {Guidetti}}, \bibinfo {author}
  {\bibfnamefont {D.}~\bibnamefont {Iñiguez}}, \bibinfo {author}
  {\bibfnamefont {A.}~\bibnamefont {Maiorano}}, \bibinfo {author}
  {\bibfnamefont {E.}~\bibnamefont {Marinari}}, \bibinfo {author}
  {\bibfnamefont {V.}~\bibnamefont {Martin-Mayor}}, \bibinfo {author}
  {\bibfnamefont {J.}~\bibnamefont {Monforte-Garcia}}, \bibinfo {author}
  {\bibfnamefont {A.}~\bibnamefont {Muñoz~Sudupe}}, \bibinfo {author}
  {\bibfnamefont {D.}~\bibnamefont {Navarro}}, \bibinfo {author} {\bibfnamefont
  {G.}~\bibnamefont {Parisi}}, \bibinfo {author} {\bibfnamefont
  {S.}~\bibnamefont {Perez-Gaviro}}, \bibinfo {author} {\bibfnamefont {J.~J.}\
  \bibnamefont {Ruiz-Lorenzo}}, \bibinfo {author} {\bibfnamefont {S.~F.}\
  \bibnamefont {Schifano}}, \bibinfo {author} {\bibfnamefont {B.}~\bibnamefont
  {Seoane}}, \bibinfo {author} {\bibfnamefont {A.}~\bibnamefont {Tarancon}},
  \bibinfo {author} {\bibfnamefont {P.}~\bibnamefont {Tellez}}, \bibinfo
  {author} {\bibfnamefont {R.}~\bibnamefont {Tripiccione}}, \ and\ \bibinfo
  {author} {\bibfnamefont {D.}~\bibnamefont {Yllanes}},\ }\href {\doibase
  10.1073/pnas.1203295109} {\bibfield  {journal} {\bibinfo  {journal}
  {Proceedings of the National Academy of Sciences}\ }\textbf {\bibinfo
  {volume} {109}},\ \bibinfo {pages} {6452} (\bibinfo {year} {2012})},\ \Eprint
  {http://arxiv.org/abs/http://www.pnas.org/content/109/17/6452.full.pdf+html}
  {http://www.pnas.org/content/109/17/6452.full.pdf+html} \BibitemShut
  {NoStop}%
\bibitem [{\citenamefont {Baity-Jesi}\ \emph {et~al.}(2014)\citenamefont
  {Baity-Jesi}, \citenamefont {Baños}, \citenamefont {Cruz}, \citenamefont
  {Fernandez}, \citenamefont {Gil-Narvion}, \citenamefont {Gordillo-Guerrero},
  \citenamefont {Iñiguez}, \citenamefont {Maiorano}, \citenamefont
  {Mantovani}, \citenamefont {Marinari}, \citenamefont {Martin-Mayor},
  \citenamefont {Monforte-Garcia}, \citenamefont {Sudupe}, \citenamefont
  {Navarro}, \citenamefont {Parisi}, \citenamefont {Perez-Gaviro},
  \citenamefont {Pivanti}, \citenamefont {Ricci-Tersenghi}, \citenamefont
  {Ruiz-Lorenzo}, \citenamefont {Schifano}, \citenamefont {Seoane},
  \citenamefont {Tarancon}, \citenamefont {Tripiccione},\ and\ \citenamefont
  {Yllanes}}]{field2}%
  \BibitemOpen
  \bibfield  {author} {\bibinfo {author} {\bibfnamefont {M.}~\bibnamefont
  {Baity-Jesi}}, \bibinfo {author} {\bibfnamefont {R.~A.}\ \bibnamefont
  {Baños}}, \bibinfo {author} {\bibfnamefont {A.}~\bibnamefont {Cruz}},
  \bibinfo {author} {\bibfnamefont {L.~A.}\ \bibnamefont {Fernandez}}, \bibinfo
  {author} {\bibfnamefont {J.~M.}\ \bibnamefont {Gil-Narvion}}, \bibinfo
  {author} {\bibfnamefont {A.}~\bibnamefont {Gordillo-Guerrero}}, \bibinfo
  {author} {\bibfnamefont {D.}~\bibnamefont {Iñiguez}}, \bibinfo {author}
  {\bibfnamefont {A.}~\bibnamefont {Maiorano}}, \bibinfo {author}
  {\bibfnamefont {F.}~\bibnamefont {Mantovani}}, \bibinfo {author}
  {\bibfnamefont {E.}~\bibnamefont {Marinari}}, \bibinfo {author}
  {\bibfnamefont {V.}~\bibnamefont {Martin-Mayor}}, \bibinfo {author}
  {\bibfnamefont {J.}~\bibnamefont {Monforte-Garcia}}, \bibinfo {author}
  {\bibfnamefont {A.~M.}\ \bibnamefont {Sudupe}}, \bibinfo {author}
  {\bibfnamefont {D.}~\bibnamefont {Navarro}}, \bibinfo {author} {\bibfnamefont
  {G.}~\bibnamefont {Parisi}}, \bibinfo {author} {\bibfnamefont
  {S.}~\bibnamefont {Perez-Gaviro}}, \bibinfo {author} {\bibfnamefont
  {M.}~\bibnamefont {Pivanti}}, \bibinfo {author} {\bibfnamefont
  {F.}~\bibnamefont {Ricci-Tersenghi}}, \bibinfo {author} {\bibfnamefont
  {J.~J.}\ \bibnamefont {Ruiz-Lorenzo}}, \bibinfo {author} {\bibfnamefont
  {S.~F.}\ \bibnamefont {Schifano}}, \bibinfo {author} {\bibfnamefont
  {B.}~\bibnamefont {Seoane}}, \bibinfo {author} {\bibfnamefont
  {A.}~\bibnamefont {Tarancon}}, \bibinfo {author} {\bibfnamefont
  {R.}~\bibnamefont {Tripiccione}}, \ and\ \bibinfo {author} {\bibfnamefont
  {D.}~\bibnamefont {Yllanes}},\ }\href
  {http://stacks.iop.org/1742-5468/2014/i=5/a=P05014} {\bibfield  {journal}
  {\bibinfo  {journal} {Journal of Statistical Mechanics: Theory and
  Experiment}\ }\textbf {\bibinfo {volume} {2014}},\ \bibinfo {pages} {P05014}
  (\bibinfo {year} {2014})}\BibitemShut {NoStop}%
\bibitem [{\citenamefont {Young}\ and\ \citenamefont {Katzgraber}(2004)}]{at}%
  \BibitemOpen
  \bibfield  {author} {\bibinfo {author} {\bibfnamefont {A.~P.}\ \bibnamefont
  {Young}}\ and\ \bibinfo {author} {\bibfnamefont {H.~G.}\ \bibnamefont
  {Katzgraber}},\ }\href {\doibase 10.1103/PhysRevLett.93.207203} {\bibfield
  {journal} {\bibinfo  {journal} {Phys. Rev. Lett.}\ }\textbf {\bibinfo
  {volume} {93}},\ \bibinfo {pages} {207203} (\bibinfo {year}
  {2004})}\BibitemShut {NoStop}%
\bibitem [{\citenamefont {Katzgraber}\ \emph {et~al.}(2009)\citenamefont
  {Katzgraber}, \citenamefont {Larson},\ and\ \citenamefont
  {Young}}]{katzgraber09}%
  \BibitemOpen
  \bibfield  {author} {\bibinfo {author} {\bibfnamefont {H.~G.}\ \bibnamefont
  {Katzgraber}}, \bibinfo {author} {\bibfnamefont {D.}~\bibnamefont {Larson}},
  \ and\ \bibinfo {author} {\bibfnamefont {A.~P.}\ \bibnamefont {Young}},\
  }\href {\doibase 10.1103/PhysRevLett.102.177205} {\bibfield  {journal}
  {\bibinfo  {journal} {Phys. Rev. Lett.}\ }\textbf {\bibinfo {volume} {102}},\
  \bibinfo {pages} {177205} (\bibinfo {year} {2009})}\BibitemShut {NoStop}%
\bibitem [{\citenamefont {Larson}\ \emph {et~al.}(2013)\citenamefont {Larson},
  \citenamefont {Katzgraber}, \citenamefont {Moore},\ and\ \citenamefont
  {Young}}]{larson13}%
  \BibitemOpen
  \bibfield  {author} {\bibinfo {author} {\bibfnamefont {D.}~\bibnamefont
  {Larson}}, \bibinfo {author} {\bibfnamefont {H.~G.}\ \bibnamefont
  {Katzgraber}}, \bibinfo {author} {\bibfnamefont {M.~A.}\ \bibnamefont
  {Moore}}, \ and\ \bibinfo {author} {\bibfnamefont {A.~P.}\ \bibnamefont
  {Young}},\ }\href {\doibase 10.1103/PhysRevB.87.024414} {\bibfield  {journal}
  {\bibinfo  {journal} {Phys. Rev. B}\ }\textbf {\bibinfo {volume} {87}},\
  \bibinfo {pages} {024414} (\bibinfo {year} {2013})}\BibitemShut {NoStop}%
\bibitem [{\citenamefont {Imry}\ and\ \citenamefont {Ma}(1975)}]{imryma}%
  \BibitemOpen
  \bibfield  {author} {\bibinfo {author} {\bibfnamefont {Y.}~\bibnamefont
  {Imry}}\ and\ \bibinfo {author} {\bibfnamefont {S.-k.}\ \bibnamefont {Ma}},\
  }\href {\doibase 10.1103/PhysRevLett.35.1399} {\bibfield  {journal} {\bibinfo
   {journal} {Phys. Rev. Lett.}\ }\textbf {\bibinfo {volume} {35}},\ \bibinfo
  {pages} {1399} (\bibinfo {year} {1975})}\BibitemShut {NoStop}%
\bibitem [{\citenamefont {Fisher}\ and\ \citenamefont {Huse}(1986)}]{droplet}%
  \BibitemOpen
  \bibfield  {author} {\bibinfo {author} {\bibfnamefont {D.~S.}\ \bibnamefont
  {Fisher}}\ and\ \bibinfo {author} {\bibfnamefont {D.~A.}\ \bibnamefont
  {Huse}},\ }\href {\doibase 10.1103/PhysRevLett.56.1601} {\bibfield  {journal}
  {\bibinfo  {journal} {Phys. Rev. Lett.}\ }\textbf {\bibinfo {volume} {56}},\
  \bibinfo {pages} {1601} (\bibinfo {year} {1986})}\BibitemShut {NoStop}%
\bibitem [{\citenamefont {Kawashima}(2000)}]{Kawashima99}%
  \BibitemOpen
  \bibfield  {author} {\bibinfo {author} {\bibfnamefont {N.}~\bibnamefont
  {Kawashima}},\ }\href {\doibase 10.1143/JPSJ.69.987} {\bibfield  {journal}
  {\bibinfo  {journal} {Journal of the Physical Society of Japan}\ }\textbf
  {\bibinfo {volume} {69}},\ \bibinfo {pages} {987} (\bibinfo {year} {2000})},\
  \Eprint {http://arxiv.org/abs/http://dx.doi.org/10.1143/JPSJ.69.987}
  {http://dx.doi.org/10.1143/JPSJ.69.987} \BibitemShut {NoStop}%
\bibitem [{\citenamefont {Kawashima}\ and\ \citenamefont
  {Aoki}(2000)}]{Kawashima00}%
  \BibitemOpen
  \bibfield  {author} {\bibinfo {author} {\bibfnamefont {N.}~\bibnamefont
  {Kawashima}}\ and\ \bibinfo {author} {\bibfnamefont {T.}~\bibnamefont
  {Aoki}},\ }\href@noop {} {\bibfield  {journal} {\bibinfo  {journal} {J. Phys.
  Soc. Jpn}\ }\textbf {\bibinfo {volume} {69}},\ \bibinfo {pages} {169}
  (\bibinfo {year} {2000})}\BibitemShut {NoStop}%
\bibitem [{\citenamefont {Hartmann}\ and\ \citenamefont
  {Moore}(2004)}]{hartmann04}%
  \BibitemOpen
  \bibfield  {author} {\bibinfo {author} {\bibfnamefont {A.~K.}\ \bibnamefont
  {Hartmann}}\ and\ \bibinfo {author} {\bibfnamefont {M.~A.}\ \bibnamefont
  {Moore}},\ }\href {\doibase 10.1103/PhysRevB.69.104409} {\bibfield  {journal}
  {\bibinfo  {journal} {Phys. Rev. B}\ }\textbf {\bibinfo {volume} {69}},\
  \bibinfo {pages} {104409} (\bibinfo {year} {2004})}\BibitemShut {NoStop}%
\bibitem [{\citenamefont {Bertele}\ and\ \citenamefont
  {Brioschi}(1972)}]{treewidth}%
  \BibitemOpen
  \bibfield  {author} {\bibinfo {author} {\bibfnamefont {U.}~\bibnamefont
  {Bertele}}\ and\ \bibinfo {author} {\bibfnamefont {F.}~\bibnamefont
  {Brioschi}},\ }\href@noop {} {\emph {\bibinfo {title} {Nonserial Dynamic
  Programming}}}\ (\bibinfo  {publisher} {Academic Press, Inc.},\ \bibinfo
  {address} {Orlando, FL, USA},\ \bibinfo {year} {1972})\BibitemShut {NoStop}%
\bibitem [{Note1()}]{Note1}%
  \BibitemOpen
  \bibinfo {note} {The median was calculated from $5280$ random instances for
  each system size and field strength. In our implementation we use a library
  \cite {BDD_library} for binary decision diagrams to store these
  configurations which we found to be significantly more efficient than using
  simple boolean tables.}\BibitemShut {Stop}%
\bibitem [{\citenamefont {De~Simone}\ \emph {et~al.}(1995)\citenamefont
  {De~Simone}, \citenamefont {Diehl}, \citenamefont {Jünger}, \citenamefont
  {Mutzel}, \citenamefont {Reinelt},\ and\ \citenamefont {Rinaldi}}]{exact}%
  \BibitemOpen
  \bibfield  {author} {\bibinfo {author} {\bibfnamefont {C.}~\bibnamefont
  {De~Simone}}, \bibinfo {author} {\bibfnamefont {M.}~\bibnamefont {Diehl}},
  \bibinfo {author} {\bibfnamefont {M.}~\bibnamefont {Jünger}}, \bibinfo
  {author} {\bibfnamefont {P.}~\bibnamefont {Mutzel}}, \bibinfo {author}
  {\bibfnamefont {G.}~\bibnamefont {Reinelt}}, \ and\ \bibinfo {author}
  {\bibfnamefont {G.}~\bibnamefont {Rinaldi}},\ }\href {\doibase
  10.1007/BF02178370} {\bibfield  {journal} {\bibinfo  {journal} {Journal of
  Statistical Physics}\ }\textbf {\bibinfo {volume} {80}},\ \bibinfo {pages}
  {487} (\bibinfo {year} {1995})}\BibitemShut {NoStop}%
\bibitem [{Note2()}]{Note2}%
  \BibitemOpen
  \bibinfo {note} {We used the spin glass server \cite {SGS} to computed the
  exact ground states for three dimensional lattices}\BibitemShut {NoStop}%
\bibitem [{\citenamefont {Denil}\ and\ \citenamefont {Freitas}()}]{chimera}%
  \BibitemOpen
  \bibfield  {author} {\bibinfo {author} {\bibfnamefont {M.}~\bibnamefont
  {Denil}}\ and\ \bibinfo {author} {\bibfnamefont {N.~D.}\ \bibnamefont
  {Freitas}},\ }\href@noop {} {\enquote {\bibinfo {title} {Toward the
  implementation of a quantum rbm},}\ }\BibitemShut {NoStop}%
\bibitem [{\citenamefont {Google}()}]{chimera_cluster}%
  \BibitemOpen
  \bibfield  {author} {\bibinfo {author} {\bibnamefont {Google}},\ }\href
  {https://plus.google.com/+QuantumAILab/posts/DymNo8DzAYi} {}\bibinfo {note}
  {\url{https://plus.google.com/+QuantumAILab/posts/DymNo8DzAYi}}\BibitemShut
  {NoStop}%
\bibitem [{\citenamefont {Swendsen}\ and\ \citenamefont {Wang}(1986)}]{PT}%
  \BibitemOpen
  \bibfield  {author} {\bibinfo {author} {\bibfnamefont {R.~H.}\ \bibnamefont
  {Swendsen}}\ and\ \bibinfo {author} {\bibfnamefont {J.-S.}\ \bibnamefont
  {Wang}},\ }\href {\doibase 10.1103/PhysRevLett.57.2607} {\bibfield  {journal}
  {\bibinfo  {journal} {Phys. Rev. Lett.}\ }\textbf {\bibinfo {volume} {57}},\
  \bibinfo {pages} {2607} (\bibinfo {year} {1986})}\BibitemShut {NoStop}%
\bibitem [{\citenamefont {Katzgraber}\ \emph {et~al.}(2014)\citenamefont
  {Katzgraber}, \citenamefont {Hamze},\ and\ \citenamefont
  {Andrist}}]{2014Katzgraber}%
  \BibitemOpen
  \bibfield  {author} {\bibinfo {author} {\bibfnamefont {H.~G.}\ \bibnamefont
  {Katzgraber}}, \bibinfo {author} {\bibfnamefont {F.}~\bibnamefont {Hamze}}, \
  and\ \bibinfo {author} {\bibfnamefont {R.~S.}\ \bibnamefont {Andrist}},\
  }\href {http://link.aps.org/doi/10.1103/PhysRevX.4.021008} {\bibfield
  {journal} {\bibinfo  {journal} {Physical Review X}\ }\textbf {\bibinfo
  {volume} {4}},\ \bibinfo {pages} {021008} (\bibinfo {year}
  {2014})}\BibitemShut {NoStop}%
\bibitem [{\citenamefont {Mitchell}\ \emph {et~al.}(1992)\citenamefont
  {Mitchell}, \citenamefont {Selman},\ and\ \citenamefont
  {Levesque}}]{Mitchell1992}%
  \BibitemOpen
  \bibfield  {author} {\bibinfo {author} {\bibfnamefont {D.}~\bibnamefont
  {Mitchell}}, \bibinfo {author} {\bibfnamefont {B.}~\bibnamefont {Selman}}, \
  and\ \bibinfo {author} {\bibfnamefont {H.}~\bibnamefont {Levesque}},\ }in\
  \href {http://dl.acm.org/citation.cfm?id=1867135.1867206} {\emph {\bibinfo
  {booktitle} {Proceedings of the Tenth National Conference on Artificial
  Intelligence}}},\ \bibinfo {series and number} {AAAI'92}\ (\bibinfo
  {publisher} {AAAI Press},\ \bibinfo {year} {1992})\ pp.\ \bibinfo {pages}
  {459--465}\BibitemShut {NoStop}%
\bibitem [{\citenamefont {Lind-Nielsen}()}]{BDD_library}%
  \BibitemOpen
  \bibfield  {author} {\bibinfo {author} {\bibfnamefont {J.}~\bibnamefont
  {Lind-Nielsen}},\ }\href
  {http://vlsicad.eecs.umich.edu/BK/Slots/cache/www.itu.dk/research/buddy/}
  {}\bibinfo {note}
  {\url{http://vlsicad.eecs.umich.edu/BK/Slots/cache/www.itu.dk/research/buddy/}}\BibitemShut
  {NoStop}%
\bibitem [{SGS()}]{SGS}%
  \BibitemOpen
  \href {http://www.informatik.uni-koeln.de/spinglass/} {}\bibinfo {note} {Spin
  Glass Server \url{http://www.informatik.uni-koeln.de/spinglass/}}\BibitemShut
  {NoStop}%
\end{thebibliography}%
\end{document}